\documentclass[%
 reprint,
superscriptaddress,
amsmath,amssymb,amsfonts,
aps,
prapplied,
floatfix,
]{revtex4-2}

\usepackage[T1]{fontenc} 
\usepackage[utf8]{inputenc}

\usepackage[english]{babel}

\usepackage{hyperref}

\usepackage{graphicx}
\usepackage{dcolumn}
\usepackage{bm}

\usepackage{siunitx}
\DeclareSIUnit\angstrom{\text {Å}}
 
\usepackage{physics}
\usepackage{chemformula} 

\begin{document}
\preprint{PRApplied-NER1048/Bréhin}

\title{Gate-voltage switching of non-reciprocal transport in oxide-based Rashba interfaces}

\author{Julien Bréhin}

\author{Luis \surname{M. Vicente Arche}}

\author{Sara Varotto}

\author{Srijani Mallik}

\affiliation{Unit\'e Mixte de Physique, CNRS, Thales, Universit\'e Paris-Saclay, Palaiseau, France}

\author{Jean-Philippe Attané}
\author{Laurent Vila}
\affiliation{University Grenoble Alpes, CEA, CNRS, Grenoble INP, SPINTEC, 38000 Grenoble, France}

\author{Agnès Barthélémy}

\affiliation{Unit\'e Mixte de Physique, CNRS, Thales, Universit\'e Paris-Saclay, Palaiseau, France}

\author{Nicolas Bergeal}
\affiliation{Laboratoire de Physique et d'Etude des Matériaux, UMR 8213 - CNRS, ESPCI-Paris, PSL, Sorbonne Université, Paris, FRANCE}

\author{Manuel Bibes}
\email{Author to whom correspondence should be addressed: manuel.bibes@cnrs-thales.fr}
\affiliation{Unit\'e Mixte de Physique, CNRS, Thales, Universit\'e Paris-Saclay, Palaiseau, France}

\begin{abstract}
The linear magnetoelectric effect (ME) is the phenomenon by which an electric field produces a magnetization. Its observation requires both time-reversal and space-inversion symmetries to be broken, as in multiferroics. While the ME effect has only been studied in insulating materials, it can actually exist in non-centrosymmetric conductors such as two-dimensional electron gases (2DEGs) with Rashba spin-orbit coupling. It is then coined the Edelstein effect (EE), by which a bias voltage -- generating a charge current -- produces a transverse spin density, i.e. a magnetization. Interestingly, 2D systems are sensitive to voltage gating, which provides an extra handle to control the EE. Here, we show that the sign of the EE in a \ch{SrTiO3} 2DEG can be controlled by a gate voltage. We propose various logic devices harnessing the dual control of the spin density by current and gate voltages and discuss the potential of our findings for gate-tunable non-reciprocal electronics.
\end{abstract}


\date{{\today}}
\maketitle

\section{Introduction}


As classical semiconductor electronics reaches its physical limits and the power consumption of computing architectures increases dramatically \cite{jones_how_2018}, new paradigms of computation are emerging, notably harnessing the electron's spin in addition to its charge \cite{dieny_opportunities_2020}. Realizing beyond-CMOS devices operating on spin requires the implementation of schemes to electrically generate, control and detect spin-encoded information. Key concepts in this endeavour are the magnetoelectric effect (ME, to generate and control a magnetization by an electric field) and spin-charge interconversion. The latter can be achieved by the direct and inverse spin Hall (SHE)\cite{sinova_spin_2015} and Edelstein effects (EE)\cite{edelstein_spin_1990}. The SHE directly generates a spin current from a charge current while the inverse SHE (ISHE) does the reciprocal operation. The EE generates a spin density (that is, a magnetization) from a charge current; the spin density can then diffuse in adjacent layers as a spin current. Reciprocally, the inverse EE (IEE) generates a charge current from an injected spin density.

Examples of spin-based logic devices include Intel's MESO transistor \cite{manipatruni_scalable_2019,vaz_voltage-based_2023} (in which the ME is used to write magnetic information and the ISHE or IEE is used to read it out) or the FESO concept \cite{noel_non-volatile_2020} (that harnesses the ferroelectric control of the IEE \cite{noel_non-volatile_2020} or the ISHE \cite{fang_tuning_2020,varotto_room-temperature_2021}).


While the ME \cite{fiebig_revival_2005} and EE \cite{edelstein_spin_1990} look \textit{a priori} quite distinct, the EE may appear as a form of ME arising in conducting systems (the ME is usually restricted to insulating polar magnets such as Cr$_2$O$_3$ or multiferroics), as was actually pointed out by Edelstein himself \cite{edelstein_spin_1990}. The EE has been reported in surface states of topological insulators \cite{he_bilinear_2018} and at Rashba interfaces including Ge(111) \cite{guillet_observation_2020} and oxide interfaces \cite{choe_gate-tunable_2019,vaz_determining_2020,lee_nonreciprocal_2021,vicentearche_spincharge_2021}. However, the sign of the EE, determining the direction of the non-equilibrium magnetization generated by the applied charge current, is intrinsic to the band structure of the material and is thus usually fixed. 

Here, by injecting an electrical current in an AlO$_{x}$/SrTiO$_3$ two-dimensional electron gas (2DEG) patterned into a Hall bar, we demonstrate the generation of a magnetization via the EE, that manifests as an additional longitudinal resistance upon applying a transverse magnetic field. 
This resistance term depends linearly on the current, indicating that the sign of the EE-generated magnetization changes when switching the current direction.
We show that it is possible to also change its sign by applying a gate voltage, which tunes the Fermi energy across the competing Rashba-split electronic band pairs of the SrTiO$_3$ (STO) 2DEG\cite{vaz_mapping_2019,trier_electric-field_2020}. This provides a unique dual control of the EE-generated magnetization, which we propose to exploit in various types of spintronic logic gate devices free from ferromagnets.

\section{Theoretical background}

\begin{figure*}[ht]
\centering
\includegraphics[width=\linewidth]{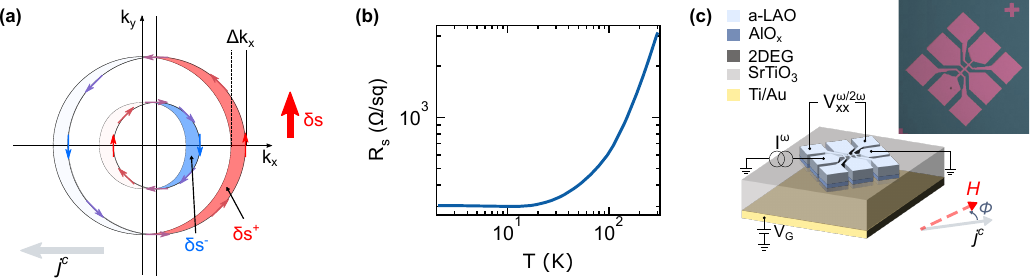}
\caption{(a) Fermi contours resulting from the direct Edelstein effect (DEE) in a simple Rashba system with spin-split parabolic bands. Upon injection of a charge current $\vb{j^c}$ along $-x$, the Fermi contours are shifted in the same direction by $\Delta k_x$, creating a net density of spins with transverse direction, and resulting in a net magnetization. (b) Temperature dependence of the sheet resistance of a device based on a-LAO/Al/STO 2DEG (LAO: LaAlO$_3$). (c) Schematic of the sample and measurement configuration used throughout the paper. The inset shows a microscope image of the actual device used hereafter.}
\label{fig:DEE_sample_configuration}
\end{figure*}

The Rashba effect \cite{bychkov_properties_1984} results from the coexistence of a broken inversion symmetry -- which is typically present at a heterostructure interface and manifests as an electric field perpendicular to that interface -- and a sizeable spin-orbit interaction (SOI).
The direct consequence is the lifting of spin-degeneracy in momentum-space, introduced by a new term $H_{R} = \frac{\alpha_R}{\hbar} (\sigma \cp p)$$ \vdot \vu{{z}}$ in the Hamiltonian, where $\hbar$ is the reduced Planck constant, $\vb*{\sigma}$ is the vector of Pauli matrices, $\vb{p}$ is the electron momentum and $\vu{z}$ is a unit vector normal to the plane of the heterointerface (and thus, parallel to the symmetry-breaking electric field) and $\alpha_R$ is a constant that quantifies the strength of the Rashba SOI and the resulting spin-splitting.
It appears from this term that the spin and momentum degrees of freedom will be locked perpendicular to each other and, in a system where the Rashba splitting occurs near the Fermi level, the Fermi surface will consist of two concentric contours with opposite chirality of the spin winding \cite{manchon_new_2015}.

In the semiclassical picture of transport, applying an electric field across the 2DEG and thus injecting a charge current \(\vb{j^c}\) along one direction, say \(\vb{-x}\), results in a shift of both inner and outer contours by the same amount \(\Delta k_x\). This generates an excess spin densities \(\delta s_{+}\) and \(\delta s_{-}\) for each contour as seen in Fig. \ref{fig:DEE_sample_configuration}a.
Since these two contours do not coincide, these two densities are not equivalent, they do not compensate each other and a finite spin density (a magnetization) transverse  (i.e. along the $y$-axis) and proportional to the charge current appears in the system.
This magneto-electric effect converting a charge current into a spin current was first described in Ref. \cite{edelstein_spin_1990} and is referred to as the direct EE or the inverse spin galvanic effect.

\begin{figure*}[ht!]
\centering
\includegraphics[width=\linewidth]{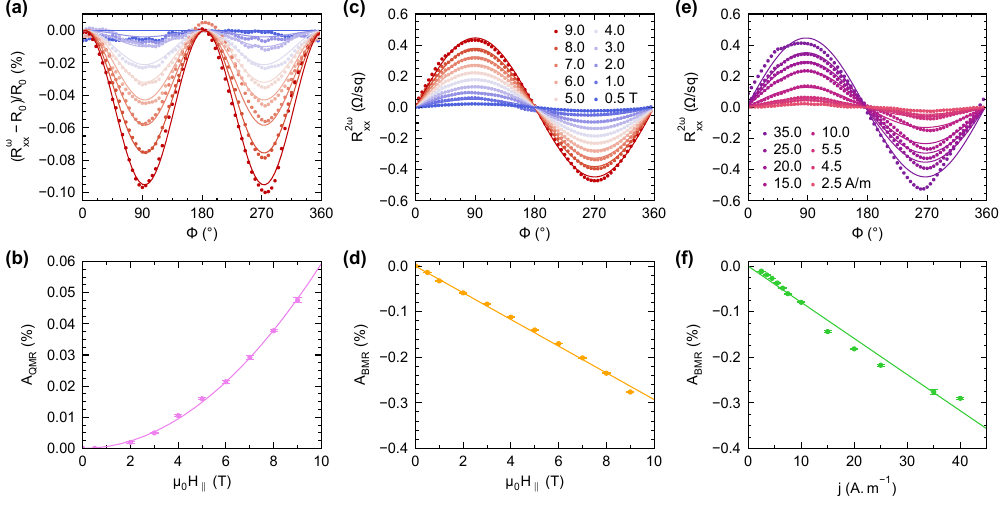}
\caption{Angle dependence of the longitudinal resistance obtained by harmonic transport with an in-plane magnetic field, at a temperature $T = \SI{2}{\kelvin}$ and gate voltage $V_g = \SI{200}{\volt}$.
(a) First harmonic signal normalized by the (zero-field) sheet resistance $R_0 = R_s (B = 0)$, for magnetic fields at \SI{0.5}{\tesla} (in blue) and increasing from \SI{2}{\tesla} up to \SI{9}{\tesla} (in red) with steps of \SI{1}{\tesla}. Continuous lines are fits to the experimental data (filled circles). 
(b) Fitted amplitudes of the first harmonic resistance signal showing a quadratic dependence with the in-plane magnetic field amplitude $\mu_0 H_\parallel$.
(c) and (e) angular dependences of the second harmonic signal $R_{xx}^{2\omega}$ at different fields (c) and current densities (e), showing unidirectional $2\pi$-periodic magnetoresistance.
(d) and (f) display the corresponding fits of their amplitude as a function of field and current, respectively. Some points on (f) at low current densities correspond to curves not shown on (e) for clarity.}
\label{fig:BMR_QMR_vs_H_j}
\end{figure*}

It is important to note that, upon reversal of the sign of the injected charge current, the net magnetization generated will change its orientation accordingly.
Thus, an external magnetic field \(\vb{B}\) applied in-plane will interact constructively or destructively with the generated magnetization depending on the sign of the current and the applied field orientation, leading to a decrease or to an increase of the sample resistance.
This unidirectional magnetoresistance term then has two extrema as a function of the angle \(\Phi\) between the current and the field, namely when they are transverse to each other, and it is proportional to both the injected current density and the applied magnetic field. As such, it is also called bilinear magnetoresistance (BMR) and its angular dependence has been found by Vaz \emph{et al.} \cite{vaz_determining_2020} to follow:
\begin{equation}
    \textup{BMR} = A_{BMR} \frac{j^c_x}{j^c} \sin{\Phi}, \label{eq:BMR}
\end{equation}
where 
\begin{equation}
    A_{BMR} = \frac{3}{2} \pi \frac{g\mu_B}{\abs{e}\hbar} \frac{\alpha_R \tau^2}{\epsilon_F} j^c B, \label{eq:BMRamplitude}
\end{equation}
in the case where one uses normalized magnetoresistance \( \textup{MR} = [R_s(B) - R_s(B=0)]/R_s(B=0) \) and with \(\Phi\) the angle between the injected charge current \(\vb{j^c}\) and the applied in-plane magnetic field \(\vb{B} = \mu_0 \vb{H_\parallel}\).
For a given set of experimental conditions, the amplitude of this effect can thus be used to estimate the Rashba coefficient \(\alpha_R\). However, it is also highly dependent on other physical parameters of the system such as the scattering time \(\tau\), the \(g\)-factor and Fermi energy \(\epsilon_F\).
Importantly, in the case of oxide 2DEGs, it is known that these three parameters can vary when applying a gate voltage, in addition to \(\alpha_R\) \cite{caviglia_tunable_2010,lesne_highly_2016,vaz_mapping_2019}. 


Concurrently to the BMR, the magnetoresistance is also modulated by an anisotropic \(\pi\)-periodic term that is independent on the current but that scales quadratically with the applied magnetic field, i.e. a quadratic magnetoresistance (QMR) term \cite{vaz_determining_2020,boudjada_anisotropic_2019}:
\begin{equation}
    \textup{QMR} = A_{QMR} \cos{2\Phi}, \label{eq:QMR}
\end{equation}
where
\begin{equation}
    A_{QMR} = \frac{3}{4} \left(\frac{g\mu_B}{\hbar}\right)^2 \tau^2 B^2. \label{eq:QMRamplitude}
\end{equation}

Since both the BMR and QMR are increasing with \(\tau^2\), taking the ratio between their amplitudes removes one system-dependent variable and simplifies the expression into a function depending mainly on universal constants and fixed experimental parameters:
\begin{equation}
    \frac{A_{BMR}}{A_{QMR}} = \frac{2\pi\hbar}{\abs{e}g\mu_B} \frac{\alpha_R}{\epsilon_F} \frac{j^c}{B}.
\end{equation}

This model provides an efficient method to extract a Rashba coefficient in different systems, and has been applied already to \ch{LAO/STO} and \ch{KTaO3} 2DEGs, yielding values consistent with the literature\cite{caviglia_tunable_2010,zhang_unusual_2019,varotto_direct_2022}.
In \ch{LAO/STO}, Vaz \emph{et al.} have shown a gate-induced modulation by a factor \(\sim 2.5\) of the Rashba coefficient extracted from the BMR\cite{vaz_determining_2020}.
In the next section, we show results for \ch{AlO_{x}/STO} 2DEGs with a carrier density typically higher than in LAO/STO \cite{vicente-arche_metal_2021} and evidence a change of sign of the BMR and thus of the effective \(\alpha_R\) by a gate voltage.

\section{Experimental details}

\subsection{Sample preparation}

To characterize the charge-to-spin current conversion, we have performed transport measurements in a 2DEG formed by depositing \SI{2}{nm} of Al on a STO substrate \cite{vicente-arche_metal_2021}.
We fabricated Hall bar devices \textit{via} a single-step lift-off process by exposing the design with the SmartPrint projection lithography setup after spin-coating a SPR 700 1.0 photoresist on the bare substrate.
Al was deposited at room temperature by dc magnetron sputtering on as received STO (001) substrates from SurfaceNet.
During Al deposition, the Ar partial pressure and the dc power were kept fixed at $5.2\times10^{-4}$ mbar and 11 W, respectively so that the deposition rate was with a rate of $\sim$ 1 \AA/s.

The Al in contact with the STO substrate oxidizes into \ch{AlO_{x}}, which reduces the STO and dopes its surface with electrons, forming the 2DEG\cite{vicente-arche_metal_2021}.
To help increasing the visibility of the devices, a layer of \SI{20}{nm} of amorphous LaAlO$_3$ (a-LAO) was then deposited by PLD at room temperature with a KrF Excimer laser at an energy of \SI{42}{\milli\joule}, before proceeding to the lift-off with acetone.

\subsection{Magnetotransport measurements} 
The samples were bonded with Al wire after the patterning process and inserted in a Quantum Design PPMS to measure their transport properties.
They show the usual metallic behaviour typical of \ch{AlO_{x}/STO} 2DEGs, with a residual resistivity ratio of \si{\sim 10} between \SIlist{300;2}{\kelvin}, as seen in Fig. \ref{fig:DEE_sample_configuration} (b).

At the lowest temperature of \SI{2}{\kelvin}, a back-gate forming step was applied: the gate voltage was ﬁrst increased to its maximum value \(V_{G,max}\) and then cycled repeatedly within the gate range of the experiment \([-V_{G,max}; +V_{G,max}]\) to suppress hysteresis and ensure the full reversibility of the measurements.
Figure \ref{fig:BMR_QMR_vs_H_j} and \ref{fig:BMR_vs_gate} were obtained from two distinct cooldowns, with \(V_{G,max} = \SI{+200}{\volt}\) and \(V_{G,max} = \SI{+150}{\volt}\), respectively.

The magnetotransport measurements were then performed by rotating the sample under an in-plane constant magnetic field while injecting an a.c. current along the main channel of the device and measuring the longitudinal voltage using Stanford Research 830 lock-in amplifiers combined with a Keithley 6221 current source and a Keithley 2400 to apply a gate voltage.
With this technique, any resistance oscillation that is linearly dependent on the current amplitude (or more broadly any voltage signal with a quadratic current dependence) will appear in the second harmonic (i.e. at twice the frequency of the injected a.c. current, \(2 \omega\)). In contrast, symmetric terms independent of the current should be found in the first harmonic, \(\omega\).

\section{Results}

Figure \ref{fig:BMR_QMR_vs_H_j}a shows the angular dependence of the sheet resistance of the device obtained from the first harmonic and normalized by the sheet resistance measured at $B = $\SI{0}{\tesla}, and for a gate voltage of $V_g = $\SI{200}{\volt}. 
Following the model from Vaz et al. \cite{vaz_determining_2020}, the data can be fitted with a cosine of $2\Phi$ as in Eq. \ref{eq:QMR}.
The amplitudes \(A_{QMR}\) obtained from such fits follow a quadratic dependence with the applied magnetic field as seen in  Fig. \ref{fig:BMR_QMR_vs_H_j}b.

Using Eq. \ref{eq:QMRamplitude} and assuming the \(g\)-factor to be equal to 2 as was reported in LAO/STO at high gate voltages \cite{caviglia_tunable_2010}, we obtain a momentum relaxation time \(\tau_m \approx\) \SI{1.6e-14}{\second} which goes up to \SI{6.4e-14}{\second} for \(g=0.5\) (as used in ref. \cite{vaz_determining_2020}).
This is in the range of values obtained from a classical transport model with magnetotransport data obtained in the same sample (see Fig S1 in Supplemental Material \cite{supplemental_material}), with \(\tau_m \approx\) \SIrange{5.5e-14}{1.2e-13}{\second} at the highest gate voltage (for a \(d_{xy}\)-only mass \(m^* = 0.6 m_e\) \cite{vaz_mapping_2019} or an estimated \(m^* = 1.3 m_e\) used in ref. \cite{vaz_determining_2020}, respectively).

In addition, as expected from the Edelstein effect, we obtain a \(2 \pi\)-periodic signal from the second harmonic signal which we fit to extract the amplitude and plot as a function of the magnetic field in Fig. \ref{fig:BMR_QMR_vs_H_j}d and as a function of the current density in Fig. \ref{fig:BMR_QMR_vs_H_j}f.
We see that this signal is mostly linear in both current density and magnetic field, hence it corresponds to the BMR term from charge-to-spin conversion.
Note that here, we chose to use the same sign convention as in Ref. \cite{vaz_determining_2020} where a negative sine function is associated to a positive amplitude and as a consequence, the positive sine function obtained at a gate voltage of \SI{200}{\volt} yields a negative BMR amplitude.

\begin{figure}[ht]
\centering
\includegraphics[width=\linewidth]{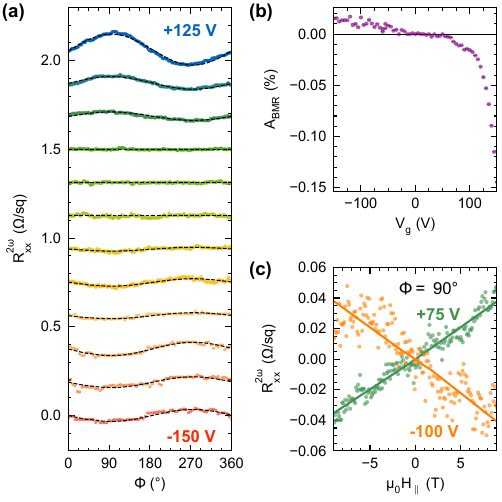}
\caption{Gate dependence of the bilinear magnetoresistance obtained from second harmonic transport measurements with \(I_{ac} = \SI{150}{\micro\ampere}\), at a temperature $T = \SI{2}{\kelvin}$.
(a) Angle-dependence of the second harmonic signal at fixed gate voltages from \SIrange{-150}{+125}{\volt}, applying \SI{9}{\tesla} in the plane of the sample. Circles represent the measured data, after removing some background contributions (see Methods), while the fits are indicated by black dashed lines. Only few curves are displayed at gate voltages separated by \SI{25}{\volt} for the sake of clarity. 
(b) BMR amplitude extracted from the fits to the angular dependences as a function of the gate voltage. (c) Sheet resistance from the second harmonic channel as the in-plane magnetic field is swept between \qtylist{+9;-9}{\tesla}, at \(V_g = \SI{-100}{\volt}\) (in orange) and \(\SI{+75}{\volt}\) (in green). The lines are a guide to the eye.}
\label{fig:BMR_vs_gate}
\end{figure}

Since the Rashba spin-orbit properties of STO 2DEGs are known to vary significantly when applying a gate voltage\cite{caviglia_tunable_2010,lesne_highly_2016,vaz_mapping_2019,vaz_determining_2020}, we have also performed similar measurements at different back-gate voltages, as shown in Fig. \ref{fig:BMR_vs_gate}.

Fig. \ref{fig:BMR_vs_gate}a shows the second-harmonic signal measured at different gate voltages varying from \qtyrange{+125}{-150}{\volt}.
It is clear that the amplitude of the BMR signal changes significantly and even switches sign, passing from positive to negative amplitude as the gate voltage is increased.
The full dependence of \(A_{BMR}\) is summarized in Fig. \ref{fig:BMR_vs_gate}b where we can identify the cross-over to be around \(V_g \approx \qtyrange{0}{20}{\volt}\).
We also measured this second-harmonic signal at a fixed angle \(\Phi = \SI{90}{\degree}\), i.e. keeping the current and field perpendicular to each other, while sweeping the magnetic field.
Fig. \ref{fig:BMR_vs_gate}c confirms the linear dependence of the BMR with the field with opposite slope between \(V_g = \SI{-100}{\volt}\) and \(\SI{+75}{\volt}\). 

\begin{figure*}[ht!]
\centering
\includegraphics[width=\linewidth]{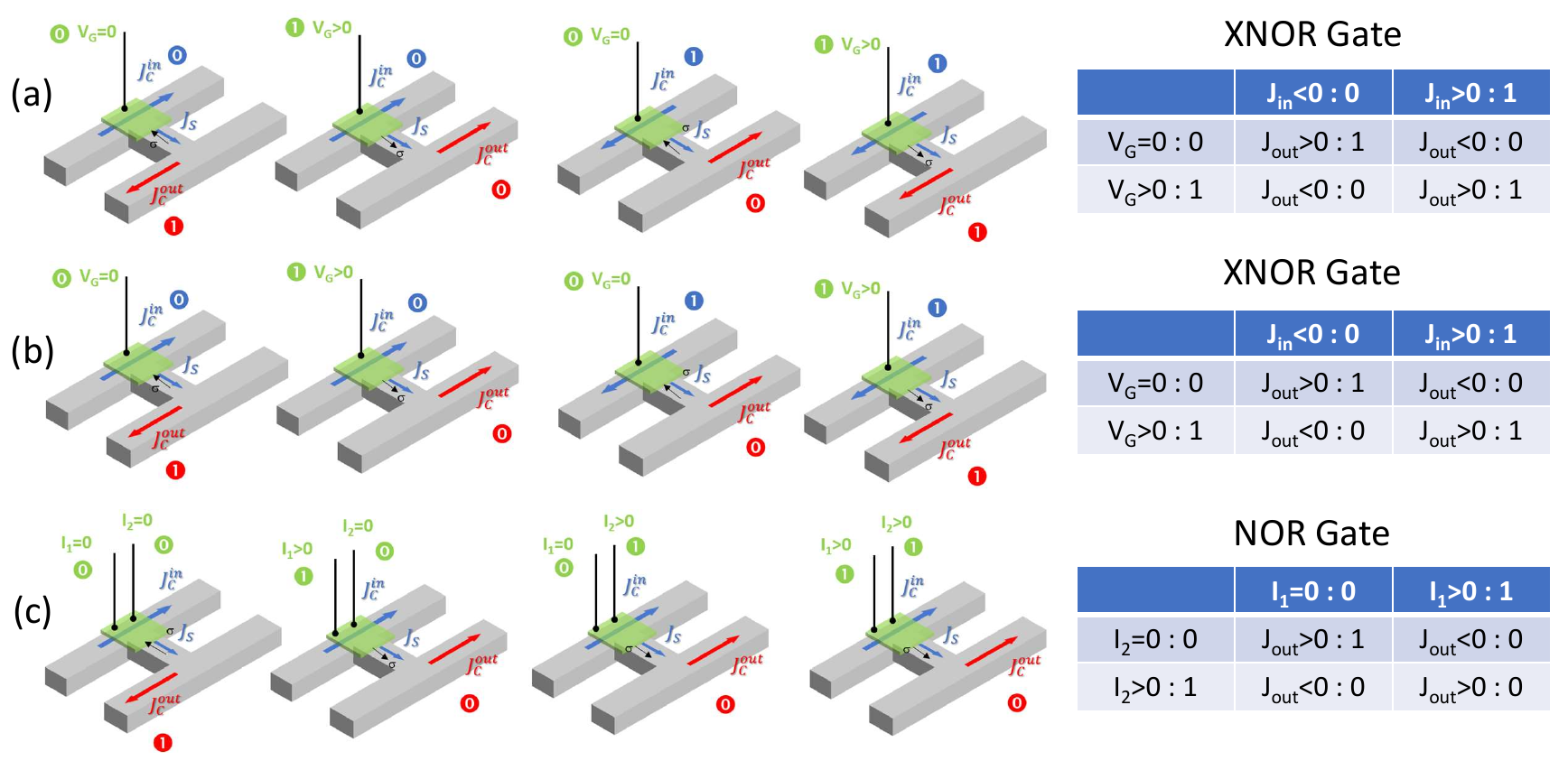}
\caption{Examples of three arms logic gate devices exploiting the direct and inverse Edelstein effects. (a) A gate voltage and the charge current sign of the first branch are used as inputs, and the sign of the current in the third branch as output, realizing the XNOR logic function. (b) Two independent gate voltages as inputs are applied at the crossing points, with the sign of the current in the third branch as output (XNOR logic gate). (c) Two charge currents are summed together and used as inputs to gate the first crossing, with the sign of the current in the third arm as the output (NOR logic gate).}
\label{fig:ISH_EE_HLogicDevice}
\end{figure*}

Checking upon Eq. \ref{eq:BMRamplitude}, we see that the BMR amplitude may depend on different systems parameters which can be affected under the application of a back-gate voltage \(V_g\).
First, variations of the \(g\)-factor have been reported from 0.5 to 2.5 in LAO/STO 2DEGs as \(V_g\) is increased towards positive values \cite{caviglia_tunable_2010}. Yet, this variation cannot explain the sign change of the BMR. In the simple parabolic band model used in ref. \cite{vaz_determining_2020}, \(A_{BMR}\) is also inversely proportional to the Fermi energy and should thus decrease in amplitude as the carrier density is increased but should remain positive. Finally, even considering the quadratic dependence of \(A_{BMR}\) with the momentum relaxation time \(\tau\), and an increase of the overall mobility as the second type of carrier (from \(d_{xz}\)/\(d_{yz}\) subbands) arises, one could explain the fast increase of the BMR amplitude in absolute value but not the sign change. 

In fact, it has been shown previously that the complex band structure of STO 2DEGs with topologically trivial and non-trivial avoided crossings and opposite contributions to the Edelstein tensor between the different subbands with \(d_{xy}\) and \(d_{xz}\)/\(d_{yz}\) character could give rise to varying amplitude and sign of the effective Rashba coefficient \(\alpha_R\) and resulting charge-spin interconversion \cite{vaz_mapping_2019,vaz_determining_2020,trier_electric-field_2020,johansson_spin_2021}.


Thus, the sign change observed in the BMR signals upon varying the applied voltage most likely reflects a change of \(\alpha_R\) as the gate voltage tunes the chemical potential of the system, populating higher states in the band-structure with \(d_{xz,yz}\) character and opposite Rashba spin texture to that of \(d_{xy}\) lower energy states.

More quantitatively, in the explored gate range and using g=0.5, within the model of Ref. \cite{vaz_determining_2020} the value of $\alpha_R$ varies from \SI{\sim 50}{\milli\electronvolt\angstrom} at large negative voltage to \SI{\sim -200}{\milli\electronvolt\angstrom} at large positive voltage. This former value is in the range of what was found in previous reports \cite{caviglia_tunable_2010}. 
However, \SI{\sim -200}{\milli\electronvolt\angstrom} clearly exceeds what is expected for STO 2DEGs. 
Using \(g\) factor values of 2 reported at high carrier densities\cite{caviglia_tunable_2010} would make this value even larger, signaling the limits of the model with single parabolic Rashba bands in this case.
Overall, these results thus point to the need for developing more elaborate models to compute the BMR from the band structure, beyond the simple one-band scenario from Ref. \cite{vaz_determining_2020}, i.e., models able to capture better the physics of this system when 3 or 4 band pairs are occupied, with a complex electronic structure involving trivial and non-trivial avoided crossings \cite{vaz_mapping_2019,johansson_spin_2021}.

\section{Discussion}

In any case, the observed BMR sign change implies that the magnetization generated via the direct Edelstein effect, when injecting a given charge current along a device main channel, changes sign as well. Since the BMR depends linearly on the current, our results thus demonstrate a dual control of the BMR (and thus of the current-generated magnetization) by both the current and the gate voltage. This offers interesting possibilities to design new spintronic devices. In a H-shaped device such as those sketched in Fig. \ref{fig:ISH_EE_HLogicDevice}, consisting of two longitudinal arms (left, input and right, output) connected by a transverse arm, applying a charge current in the left arm generates a transverse magnetization able to diffuse as a spin current in the transverse arm  over a certain length \(\lambda_s\) in the absence of any net charge displacement.
It is then possible to use the IEE at the intersecton of the transverse and right arms to convert this spin current back to a charge current whose sign will depend on \(\alpha_R\) and on the direction of the spins generated by the EE.

In Fig. \ref{fig:ISH_EE_HLogicDevice}, we present devices utilizing the direct and inverse Edelstein effects, combined with a gate electric field tuning of the 2DEG band filling, allowing a modulation of the spin-charge interconversion sign and efficiency accordingly.
Here, we have demonstrated that the BMR sign can be modulated by a back gate, and for the devices, we propose to use a top gate as it was also shown to efficiently modulate transport properties in STO 2DEGs (see e.g. \cite{hurand_field-effect_2015,jouan_multiband_2022}) and would be advantageous to integrate and scale down such devices while tuning the required voltages given some adequate material engineering.
In Fig. \ref{fig:ISH_EE_HLogicDevice}a, a top gate is thus placed at the intersection between the input and transverse arms to modulate the direction of the spins generated by the direct EE. Depending on their direction, the charge current generated by the IEE at the intersection of the transverse and output arms will be positive or negative.
Since changing the sign of the charge current in the input arm also reverses the direction of the generated spins and thus of the sign of the output current, we can exploit this dual control to perform logic operations, with the sign of the input charge current \(J_C^{in}\) and the sign of the gate voltage \(V_G\) as inputs and the sign of the output current as the ouput.
Compiling the different input and output logic states in a truth table shows that this type of device realizes an exclusive-inverted-OR (XNOR) logic function and could be chained with other similar devices. 

In Fig. \ref{fig:ISH_EE_HLogicDevice}(b), a similar logic operation is achieved but now by having two independent top gates, one controlling the sign of charge-spin conversion at the intersection of the input and transverse arms (V$_{G1}$) and another one controlling the sign of charge-spin conversion at the intersection of the transverse and output arms (V$_{G2}$).
Finally, Fig. \ref{fig:ISH_EE_HLogicDevice}(c) sketches another device with just one top gate, in which the two inputs are currents $I_1$ and $I_2$ accumulating or depleting charges at the top gate.
In absence of any current, the logic gate is in its \textit{high} state as per the truth table in Fig \ref{fig:ISH_EE_HLogicDevice}(a).
As soon as one of these two currents becomes greater than zero, an effective electric field develops across the gate capacitor, so that the output current of the device $J_{out}$ becomes negative and the logic state becomes zero.
The same results is obtained if both currents $I_1$ and $I_2$ are above zero, completing the NOR universal logic function.

\section{Conclusion}

Our work expands the topic of nonreciprocal transport phenomena\cite{tokura_nonreciprocal_2018} that is pervading the realm of quantum materials. Nonreciprocal transport has been reported in non-magnetic topological insulators\cite{he_bilinear_2018,fu_bilinear_2022}, magnetic topological insulators \cite{yasuda_large_2016,zhang_controlled_2022}, non-centrosymmetric superconductors\cite{wakatsuki_nonreciprocal_2017}, ferroelectric Rashba semiconductors \cite{li_nonreciprocal_2021} and Rashba interfaces\cite{choe_gate-tunable_2019,guillet_observation_2020,vaz_determining_2020,vicentearche_spincharge_2021}. While the mechanism may be different depending on the materials systems, the possibility to tune the nonreciprocal transport response by a gate voltage in amplitude \emph{and} sign has been reported in a few systems\cite{fu_bilinear_2022,zhang_controlled_2022} which could ultimately bring the proposed devices to higher operating temperatures. 
The theoretical efficiency of such devices is discussed in the Supplemental Material \cite{[{See }][{ for a discussion on the Rashba vs. Fermi energies and on the device output/input efficiency.}]supplemental_material} (see also references \cite{cavigliaElectricFieldControl2008,joshuaUniversalCriticalDensity2012,biscarasLimitElectrostaticDoping2015,wangRoomTemperatureGiantChargetoSpin2017,shashankRoomTemperatureCharge2023,varignonUnexpectedAntagonismFerroelectricity2023} therein).
This body of results offers exciting possibilities to design spin–orbitronics applications operating at low power, from logic devices such as those proposed here to spin circulators, mixers, isolators, etc, in analogy with nonreciprocal devices operating with photons or magnons\cite{tokura_nonreciprocal_2018}.

\begin{acknowledgements}

This project received funding from the ERC Advanced grant “FRESCO” $\#$833973, from the European Commission under the H2020 FETPROACT Grant TOCHA (824140), and the French Research Agency (ANR) as part of the projects CONTRABASS (ANR-20-CE24-0023) and QUANTOP (ANR-19-CE47-0006-02).

\end{acknowledgements}

\section*{Author contributions}

JB prepared the samples with help from LMVA. JB collected and analysed the data with SV, LVMA, SM, NB and MB. MB conceptualized and led the study with NB and help from AB, LV and JPA. JB and MB wrote the paper with inputs from all authors.


\begin{thebibliography}{42}%
\makeatletter
\providecommand \@ifxundefined [1]{%
 \@ifx{#1\undefined}
}%
\providecommand \@ifnum [1]{%
 \ifnum #1\expandafter \@firstoftwo
 \else \expandafter \@secondoftwo
 \fi
}%
\providecommand \@ifx [1]{%
 \ifx #1\expandafter \@firstoftwo
 \else \expandafter \@secondoftwo
 \fi
}%
\providecommand \natexlab [1]{#1}%
\providecommand \enquote  [1]{``#1''}%
\providecommand \bibnamefont  [1]{#1}%
\providecommand \bibfnamefont [1]{#1}%
\providecommand \citenamefont [1]{#1}%
\providecommand \href@noop [0]{\@secondoftwo}%
\providecommand \href [0]{\begingroup \@sanitize@url \@href}%
\providecommand \@href[1]{\@@startlink{#1}\@@href}%
\providecommand \@@href[1]{\endgroup#1\@@endlink}%
\providecommand \@sanitize@url [0]{\catcode `\\12\catcode `\$12\catcode
  `\&12\catcode `\#12\catcode `\^12\catcode `\_12\catcode `\%12\relax}%
\providecommand \@@startlink[1]{}%
\providecommand \@@endlink[0]{}%
\providecommand \url  [0]{\begingroup\@sanitize@url \@url }%
\providecommand \@url [1]{\endgroup\@href {#1}{\urlprefix }}%
\providecommand \urlprefix  [0]{URL }%
\providecommand \Eprint [0]{\href }%
\providecommand \doibase [0]{https://doi.org/}%
\providecommand \selectlanguage [0]{\@gobble}%
\providecommand \bibinfo  [0]{\@secondoftwo}%
\providecommand \bibfield  [0]{\@secondoftwo}%
\providecommand \translation [1]{[#1]}%
\providecommand \BibitemOpen [0]{}%
\providecommand \bibitemStop [0]{}%
\providecommand \bibitemNoStop [0]{.\EOS\space}%
\providecommand \EOS [0]{\spacefactor3000\relax}%
\providecommand \BibitemShut  [1]{\csname bibitem#1\endcsname}%
\let\auto@bib@innerbib\@empty
\bibitem [{\citenamefont {Jones}(2018)}]{jones_how_2018}%
  \BibitemOpen
  \bibfield  {author} {\bibinfo {author} {\bibfnamefont {N.}~\bibnamefont
  {Jones}},\ }\bibfield  {title} {{\selectlanguage {english}\bibinfo {title}
  {How to stop data centres from gobbling up the world’s electricity}},\
  }\href@noop {} {\bibfield  {journal} {\bibinfo  {journal} {Nature}\ }\textbf
  {\bibinfo {volume} {561}},\ \bibinfo {pages} {163} (\bibinfo {year}
  {2018})}\BibitemShut {NoStop}%
\bibitem [{\citenamefont {Dieny}\ \emph {et~al.}(2020)\citenamefont {Dieny},
  \citenamefont {Prejbeanu}, \citenamefont {Garello}, \citenamefont
  {Gambardella}, \citenamefont {Freitas}, \citenamefont {Lehndorff},
  \citenamefont {Raberg}, \citenamefont {Ebels}, \citenamefont {Demokritov},
  \citenamefont {Akerman}, \citenamefont {Deac}, \citenamefont {Pirro},
  \citenamefont {Adelmann}, \citenamefont {Anane}, \citenamefont {Chumak},
  \citenamefont {Hirohata}, \citenamefont {Mangin}, \citenamefont {Valenzuela},
  \citenamefont {Onbaşlı}, \citenamefont {d’Aquino}, \citenamefont
  {Prenat}, \citenamefont {Finocchio}, \citenamefont {Lopez-Diaz},
  \citenamefont {Chantrell}, \citenamefont {Chubykalo-Fesenko},\ and\
  \citenamefont {Bortolotti}}]{dieny_opportunities_2020}%
  \BibitemOpen
  \bibfield  {author} {\bibinfo {author} {\bibfnamefont {B.}~\bibnamefont
  {Dieny}}, \bibinfo {author} {\bibfnamefont {I.~L.}\ \bibnamefont
  {Prejbeanu}}, \bibinfo {author} {\bibfnamefont {K.}~\bibnamefont {Garello}},
  \bibinfo {author} {\bibfnamefont {P.}~\bibnamefont {Gambardella}}, \bibinfo
  {author} {\bibfnamefont {P.}~\bibnamefont {Freitas}}, \bibinfo {author}
  {\bibfnamefont {R.}~\bibnamefont {Lehndorff}}, \bibinfo {author}
  {\bibfnamefont {W.}~\bibnamefont {Raberg}}, \bibinfo {author} {\bibfnamefont
  {U.}~\bibnamefont {Ebels}}, \bibinfo {author} {\bibfnamefont {S.~O.}\
  \bibnamefont {Demokritov}}, \bibinfo {author} {\bibfnamefont
  {J.}~\bibnamefont {Akerman}}, \bibinfo {author} {\bibfnamefont
  {A.}~\bibnamefont {Deac}}, \bibinfo {author} {\bibfnamefont {P.}~\bibnamefont
  {Pirro}}, \bibinfo {author} {\bibfnamefont {C.}~\bibnamefont {Adelmann}},
  \bibinfo {author} {\bibfnamefont {A.}~\bibnamefont {Anane}}, \bibinfo
  {author} {\bibfnamefont {A.~V.}\ \bibnamefont {Chumak}}, \bibinfo {author}
  {\bibfnamefont {A.}~\bibnamefont {Hirohata}}, \bibinfo {author}
  {\bibfnamefont {S.}~\bibnamefont {Mangin}}, \bibinfo {author} {\bibfnamefont
  {S.~O.}\ \bibnamefont {Valenzuela}}, \bibinfo {author} {\bibfnamefont
  {M.~C.}\ \bibnamefont {Onbaşlı}}, \bibinfo {author} {\bibfnamefont
  {M.}~\bibnamefont {d’Aquino}}, \bibinfo {author} {\bibfnamefont
  {G.}~\bibnamefont {Prenat}}, \bibinfo {author} {\bibfnamefont
  {G.}~\bibnamefont {Finocchio}}, \bibinfo {author} {\bibfnamefont
  {L.}~\bibnamefont {Lopez-Diaz}}, \bibinfo {author} {\bibfnamefont
  {R.}~\bibnamefont {Chantrell}}, \bibinfo {author} {\bibfnamefont
  {O.}~\bibnamefont {Chubykalo-Fesenko}},\ and\ \bibinfo {author}
  {\bibfnamefont {P.}~\bibnamefont {Bortolotti}},\ }\bibfield  {title}
  {{\selectlanguage {english}\bibinfo {title} {Opportunities and challenges for
  spintronics in the microelectronics industry}},\ }\href
  {https://doi.org/10.1038/s41928-020-0461-5} {\bibfield  {journal} {\bibinfo
  {journal} {Nature Electronics}\ }\textbf {\bibinfo {volume} {3}},\ \bibinfo
  {pages} {446} (\bibinfo {year} {2020})}\BibitemShut {NoStop}%
\bibitem [{\citenamefont {Sinova}\ \emph {et~al.}(2015)\citenamefont {Sinova},
  \citenamefont {Valenzuela}, \citenamefont {Wunderlich}, \citenamefont
  {Back},\ and\ \citenamefont {Jungwirth}}]{sinova_spin_2015}%
  \BibitemOpen
  \bibfield  {author} {\bibinfo {author} {\bibfnamefont {J.}~\bibnamefont
  {Sinova}}, \bibinfo {author} {\bibfnamefont {S.~O.}\ \bibnamefont
  {Valenzuela}}, \bibinfo {author} {\bibfnamefont {J.}~\bibnamefont
  {Wunderlich}}, \bibinfo {author} {\bibfnamefont {C.}~\bibnamefont {Back}},\
  and\ \bibinfo {author} {\bibfnamefont {T.}~\bibnamefont {Jungwirth}},\
  }\bibfield  {title} {{\selectlanguage {english}\bibinfo {title} {Spin {Hall}
  effects}},\ }\href {https://doi.org/10.1103/RevModPhys.87.1213} {\bibfield
  {journal} {\bibinfo  {journal} {Reviews of Modern Physics}\ }\textbf
  {\bibinfo {volume} {87}},\ \bibinfo {pages} {1213} (\bibinfo {year}
  {2015})}\BibitemShut {NoStop}%
\bibitem [{\citenamefont {Edelstein}(1990)}]{edelstein_spin_1990}%
  \BibitemOpen
  \bibfield  {author} {\bibinfo {author} {\bibfnamefont {V.}~\bibnamefont
  {Edelstein}},\ }\bibfield  {title} {{\selectlanguage {english}\bibinfo
  {title} {Spin polarization of conduction electrons induced by electric
  current in two-dimensional asymmetric electron systems}},\ }\href
  {https://doi.org/10.1016/0038-1098(90)90963-C} {\bibfield  {journal}
  {\bibinfo  {journal} {Solid State Communications}\ }\textbf {\bibinfo
  {volume} {73}},\ \bibinfo {pages} {233} (\bibinfo {year} {1990})}\BibitemShut
  {NoStop}%
\bibitem [{\citenamefont {Manipatruni}\ \emph {et~al.}(2019)\citenamefont
  {Manipatruni}, \citenamefont {Nikonov}, \citenamefont {Lin}, \citenamefont
  {Gosavi}, \citenamefont {Liu}, \citenamefont {Prasad}, \citenamefont {Huang},
  \citenamefont {Bonturim}, \citenamefont {Ramesh},\ and\ \citenamefont
  {Young}}]{manipatruni_scalable_2019}%
  \BibitemOpen
  \bibfield  {author} {\bibinfo {author} {\bibfnamefont {S.}~\bibnamefont
  {Manipatruni}}, \bibinfo {author} {\bibfnamefont {D.~E.}\ \bibnamefont
  {Nikonov}}, \bibinfo {author} {\bibfnamefont {C.-C.}\ \bibnamefont {Lin}},
  \bibinfo {author} {\bibfnamefont {T.~A.}\ \bibnamefont {Gosavi}}, \bibinfo
  {author} {\bibfnamefont {H.}~\bibnamefont {Liu}}, \bibinfo {author}
  {\bibfnamefont {B.}~\bibnamefont {Prasad}}, \bibinfo {author} {\bibfnamefont
  {Y.-L.}\ \bibnamefont {Huang}}, \bibinfo {author} {\bibfnamefont
  {E.}~\bibnamefont {Bonturim}}, \bibinfo {author} {\bibfnamefont
  {R.}~\bibnamefont {Ramesh}},\ and\ \bibinfo {author} {\bibfnamefont {I.~A.}\
  \bibnamefont {Young}},\ }\bibfield  {title} {{\selectlanguage
  {english}\bibinfo {title} {Scalable energy-efficient magnetoelectric
  spin–orbit logic}},\ }\href {https://doi.org/10.1038/s41586-018-0770-2}
  {\bibfield  {journal} {\bibinfo  {journal} {Nature}\ }\textbf {\bibinfo
  {volume} {565}},\ \bibinfo {pages} {35} (\bibinfo {year} {2019})}\BibitemShut
  {NoStop}%
\bibitem [{\citenamefont {Vaz}\ \emph {et~al.}(2023)\citenamefont {Vaz},
  \citenamefont {Lin}, \citenamefont {Plombon}, \citenamefont {Choi},
  \citenamefont {Groen}, \citenamefont {Arango}, \citenamefont {Chuvilin},
  \citenamefont {Hueso}, \citenamefont {Nikonov}, \citenamefont {Li},
  \citenamefont {Clendenning}, \citenamefont {Gosavi}, \citenamefont {Huang},
  \citenamefont {Prasad}, \citenamefont {Ramesh}, \citenamefont {Vecchiola},
  \citenamefont {Bibes}, \citenamefont {Bouzehouane}, \citenamefont {Fusil},
  \citenamefont {Garcia}, \citenamefont {Young},\ and\ \citenamefont
  {Casanova}}]{vaz_voltage-based_2023}%
  \BibitemOpen
  \bibfield  {author} {\bibinfo {author} {\bibfnamefont {D.~C.}\ \bibnamefont
  {Vaz}}, \bibinfo {author} {\bibfnamefont {C.-C.}\ \bibnamefont {Lin}},
  \bibinfo {author} {\bibfnamefont {J.}~\bibnamefont {Plombon}}, \bibinfo
  {author} {\bibfnamefont {W.~Y.}\ \bibnamefont {Choi}}, \bibinfo {author}
  {\bibfnamefont {I.}~\bibnamefont {Groen}}, \bibinfo {author} {\bibfnamefont
  {C.}~\bibnamefont {Arango}}, \bibinfo {author} {\bibfnamefont
  {A.}~\bibnamefont {Chuvilin}}, \bibinfo {author} {\bibfnamefont {L.~E.}\
  \bibnamefont {Hueso}}, \bibinfo {author} {\bibfnamefont {D.~E.}\ \bibnamefont
  {Nikonov}}, \bibinfo {author} {\bibfnamefont {H.}~\bibnamefont {Li}},
  \bibinfo {author} {\bibfnamefont {S.~B.}\ \bibnamefont {Clendenning}},
  \bibinfo {author} {\bibfnamefont {T.~A.}\ \bibnamefont {Gosavi}}, \bibinfo
  {author} {\bibfnamefont {Y.-L.}\ \bibnamefont {Huang}}, \bibinfo {author}
  {\bibfnamefont {B.}~\bibnamefont {Prasad}}, \bibinfo {author} {\bibfnamefont
  {R.}~\bibnamefont {Ramesh}}, \bibinfo {author} {\bibfnamefont
  {A.}~\bibnamefont {Vecchiola}}, \bibinfo {author} {\bibfnamefont
  {M.}~\bibnamefont {Bibes}}, \bibinfo {author} {\bibfnamefont
  {K.}~\bibnamefont {Bouzehouane}}, \bibinfo {author} {\bibfnamefont
  {S.}~\bibnamefont {Fusil}}, \bibinfo {author} {\bibfnamefont
  {V.}~\bibnamefont {Garcia}}, \bibinfo {author} {\bibfnamefont {I.~A.}\
  \bibnamefont {Young}},\ and\ \bibinfo {author} {\bibfnamefont
  {F.}~\bibnamefont {Casanova}},\ }\bibfield  {title} {{\selectlanguage
  {english}\bibinfo {title} {Voltage-based magnetization switching and reading
  in magnetoelectric spin-orbit nanodevices}},\ }\href@noop {} {\bibfield
  {journal} {\bibinfo  {journal} {ArXiv}\  \bibinfo {pages} {2302.12162}}
  (\bibinfo {year} {2023})}\BibitemShut {NoStop}%
\bibitem [{\citenamefont {Noël}\ \emph {et~al.}(2020)\citenamefont {Noël},
  \citenamefont {Trier}, \citenamefont {Vicente~Arche}, \citenamefont
  {Bréhin}, \citenamefont {Vaz}, \citenamefont {Garcia}, \citenamefont
  {Fusil}, \citenamefont {Barthélémy}, \citenamefont {Vila}, \citenamefont
  {Bibes},\ and\ \citenamefont {Attané}}]{noel_non-volatile_2020}%
  \BibitemOpen
  \bibfield  {author} {\bibinfo {author} {\bibfnamefont {P.}~\bibnamefont
  {Noël}}, \bibinfo {author} {\bibfnamefont {F.}~\bibnamefont {Trier}},
  \bibinfo {author} {\bibfnamefont {L.~M.}\ \bibnamefont {Vicente~Arche}},
  \bibinfo {author} {\bibfnamefont {J.}~\bibnamefont {Bréhin}}, \bibinfo
  {author} {\bibfnamefont {D.~C.}\ \bibnamefont {Vaz}}, \bibinfo {author}
  {\bibfnamefont {V.}~\bibnamefont {Garcia}}, \bibinfo {author} {\bibfnamefont
  {S.}~\bibnamefont {Fusil}}, \bibinfo {author} {\bibfnamefont
  {A.}~\bibnamefont {Barthélémy}}, \bibinfo {author} {\bibfnamefont
  {L.}~\bibnamefont {Vila}}, \bibinfo {author} {\bibfnamefont {M.}~\bibnamefont
  {Bibes}},\ and\ \bibinfo {author} {\bibfnamefont {J.-P.}\ \bibnamefont
  {Attané}},\ }\bibfield  {title} {{\selectlanguage {english}\bibinfo {title}
  {Non-volatile electric control of spin–charge conversion in a {SrTiO$_3$}
  {Rashba} system}},\ }\href {https://doi.org/10.1038/s41586-020-2197-9}
  {\bibfield  {journal} {\bibinfo  {journal} {Nature}\ }\textbf {\bibinfo
  {volume} {580}},\ \bibinfo {pages} {483} (\bibinfo {year}
  {2020})}\BibitemShut {NoStop}%
\bibitem [{\citenamefont {Fang}\ \emph {et~al.}(2020)\citenamefont {Fang},
  \citenamefont {Wang}, \citenamefont {Wang}, \citenamefont {Hou},
  \citenamefont {Vetter}, \citenamefont {Kou}, \citenamefont {Yang},
  \citenamefont {Yin}, \citenamefont {Xiao}, \citenamefont {Li}, \citenamefont
  {Jiang}, \citenamefont {Lee}, \citenamefont {Zhang}, \citenamefont {Wu},
  \citenamefont {Xu}, \citenamefont {Sun},\ and\ \citenamefont
  {Shen}}]{fang_tuning_2020}%
  \BibitemOpen
  \bibfield  {author} {\bibinfo {author} {\bibfnamefont {M.}~\bibnamefont
  {Fang}}, \bibinfo {author} {\bibfnamefont {Y.}~\bibnamefont {Wang}}, \bibinfo
  {author} {\bibfnamefont {H.}~\bibnamefont {Wang}}, \bibinfo {author}
  {\bibfnamefont {Y.}~\bibnamefont {Hou}}, \bibinfo {author} {\bibfnamefont
  {E.}~\bibnamefont {Vetter}}, \bibinfo {author} {\bibfnamefont
  {Y.}~\bibnamefont {Kou}}, \bibinfo {author} {\bibfnamefont {W.}~\bibnamefont
  {Yang}}, \bibinfo {author} {\bibfnamefont {L.}~\bibnamefont {Yin}}, \bibinfo
  {author} {\bibfnamefont {Z.}~\bibnamefont {Xiao}}, \bibinfo {author}
  {\bibfnamefont {Z.}~\bibnamefont {Li}}, \bibinfo {author} {\bibfnamefont
  {L.}~\bibnamefont {Jiang}}, \bibinfo {author} {\bibfnamefont {H.~N.}\
  \bibnamefont {Lee}}, \bibinfo {author} {\bibfnamefont {S.}~\bibnamefont
  {Zhang}}, \bibinfo {author} {\bibfnamefont {R.}~\bibnamefont {Wu}}, \bibinfo
  {author} {\bibfnamefont {X.}~\bibnamefont {Xu}}, \bibinfo {author}
  {\bibfnamefont {D.}~\bibnamefont {Sun}},\ and\ \bibinfo {author}
  {\bibfnamefont {J.}~\bibnamefont {Shen}},\ }\bibfield  {title}
  {{\selectlanguage {english}\bibinfo {title} {Tuning the interfacial
  spin-orbit coupling with ferroelectricity}},\ }\href
  {https://doi.org/10.1038/s41467-020-16401-7} {\bibfield  {journal} {\bibinfo
  {journal} {Nature Communications}\ }\textbf {\bibinfo {volume} {11}},\
  \bibinfo {pages} {2627} (\bibinfo {year} {2020})}\BibitemShut {NoStop}%
\bibitem [{\citenamefont {Varotto}\ \emph {et~al.}(2021)\citenamefont
  {Varotto}, \citenamefont {Nessi}, \citenamefont {Cecchi}, \citenamefont
  {Sławińska}, \citenamefont {Noël}, \citenamefont {Petrò}, \citenamefont
  {Fagiani}, \citenamefont {Novati}, \citenamefont {Cantoni}, \citenamefont
  {Petti}, \citenamefont {Albisetti}, \citenamefont {Costa}, \citenamefont
  {Calarco}, \citenamefont {Buongiorno~Nardelli}, \citenamefont {Bibes},
  \citenamefont {Picozzi}, \citenamefont {Attané}, \citenamefont {Vila},
  \citenamefont {Bertacco},\ and\ \citenamefont
  {Rinaldi}}]{varotto_room-temperature_2021}%
  \BibitemOpen
  \bibfield  {author} {\bibinfo {author} {\bibfnamefont {S.}~\bibnamefont
  {Varotto}}, \bibinfo {author} {\bibfnamefont {L.}~\bibnamefont {Nessi}},
  \bibinfo {author} {\bibfnamefont {S.}~\bibnamefont {Cecchi}}, \bibinfo
  {author} {\bibfnamefont {J.}~\bibnamefont {Sławińska}}, \bibinfo {author}
  {\bibfnamefont {P.}~\bibnamefont {Noël}}, \bibinfo {author} {\bibfnamefont
  {S.}~\bibnamefont {Petrò}}, \bibinfo {author} {\bibfnamefont
  {F.}~\bibnamefont {Fagiani}}, \bibinfo {author} {\bibfnamefont
  {A.}~\bibnamefont {Novati}}, \bibinfo {author} {\bibfnamefont
  {M.}~\bibnamefont {Cantoni}}, \bibinfo {author} {\bibfnamefont
  {D.}~\bibnamefont {Petti}}, \bibinfo {author} {\bibfnamefont
  {E.}~\bibnamefont {Albisetti}}, \bibinfo {author} {\bibfnamefont
  {M.}~\bibnamefont {Costa}}, \bibinfo {author} {\bibfnamefont
  {R.}~\bibnamefont {Calarco}}, \bibinfo {author} {\bibfnamefont
  {M.}~\bibnamefont {Buongiorno~Nardelli}}, \bibinfo {author} {\bibfnamefont
  {M.}~\bibnamefont {Bibes}}, \bibinfo {author} {\bibfnamefont
  {S.}~\bibnamefont {Picozzi}}, \bibinfo {author} {\bibfnamefont {J.-P.}\
  \bibnamefont {Attané}}, \bibinfo {author} {\bibfnamefont {L.}~\bibnamefont
  {Vila}}, \bibinfo {author} {\bibfnamefont {R.}~\bibnamefont {Bertacco}},\
  and\ \bibinfo {author} {\bibfnamefont {C.}~\bibnamefont {Rinaldi}},\
  }\bibfield  {title} {{\selectlanguage {english}\bibinfo {title}
  {Room-temperature ferroelectric switching of spin-to-charge conversion in
  germanium telluride}},\ }\href {https://doi.org/10.1038/s41928-021-00653-2}
  {\bibfield  {journal} {\bibinfo  {journal} {Nature Electronics}\ }\textbf
  {\bibinfo {volume} {4}},\ \bibinfo {pages} {740} (\bibinfo {year}
  {2021})}\BibitemShut {NoStop}%
\bibitem [{\citenamefont {Fiebig}(2005)}]{fiebig_revival_2005}%
  \BibitemOpen
  \bibfield  {author} {\bibinfo {author} {\bibfnamefont {M.}~\bibnamefont
  {Fiebig}},\ }\bibfield  {title} {{\selectlanguage {english}\bibinfo {title}
  {Revival of the magnetoelectric effect}},\ }\href
  {https://doi.org/10.1088/0022-3727/38/8/R01} {\bibfield  {journal} {\bibinfo
  {journal} {Journal of Physics D: Applied Physics}\ }\textbf {\bibinfo
  {volume} {38}},\ \bibinfo {pages} {R123} (\bibinfo {year}
  {2005})}\BibitemShut {NoStop}%
\bibitem [{\citenamefont {He}\ \emph {et~al.}(2018)\citenamefont {He},
  \citenamefont {Zhang}, \citenamefont {Zhu}, \citenamefont {Liu},
  \citenamefont {Wang}, \citenamefont {Yu}, \citenamefont {Vignale},\ and\
  \citenamefont {Yang}}]{he_bilinear_2018}%
  \BibitemOpen
  \bibfield  {author} {\bibinfo {author} {\bibfnamefont {P.}~\bibnamefont
  {He}}, \bibinfo {author} {\bibfnamefont {S.~S.-L.}\ \bibnamefont {Zhang}},
  \bibinfo {author} {\bibfnamefont {D.}~\bibnamefont {Zhu}}, \bibinfo {author}
  {\bibfnamefont {Y.}~\bibnamefont {Liu}}, \bibinfo {author} {\bibfnamefont
  {Y.}~\bibnamefont {Wang}}, \bibinfo {author} {\bibfnamefont {J.}~\bibnamefont
  {Yu}}, \bibinfo {author} {\bibfnamefont {G.}~\bibnamefont {Vignale}},\ and\
  \bibinfo {author} {\bibfnamefont {H.}~\bibnamefont {Yang}},\ }\bibfield
  {title} {{\selectlanguage {english}\bibinfo {title} {Bilinear magnetoelectric
  resistance as a probe of three-dimensional spin texture in topological
  surface states}},\ }\href {https://doi.org/10.1038/s41567-017-0039-y}
  {\bibfield  {journal} {\bibinfo  {journal} {Nature Physics}\ }\textbf
  {\bibinfo {volume} {14}},\ \bibinfo {pages} {495} (\bibinfo {year}
  {2018})}\BibitemShut {NoStop}%
\bibitem [{\citenamefont {Guillet}\ \emph {et~al.}(2020)\citenamefont
  {Guillet}, \citenamefont {Zucchetti}, \citenamefont {Barbedienne},
  \citenamefont {Marty}, \citenamefont {Isella}, \citenamefont {Cagnon},
  \citenamefont {Vergnaud}, \citenamefont {Jaffrès}, \citenamefont {Reyren},
  \citenamefont {George}, \citenamefont {Fert},\ and\ \citenamefont
  {Jamet}}]{guillet_observation_2020}%
  \BibitemOpen
  \bibfield  {author} {\bibinfo {author} {\bibfnamefont {T.}~\bibnamefont
  {Guillet}}, \bibinfo {author} {\bibfnamefont {C.}~\bibnamefont {Zucchetti}},
  \bibinfo {author} {\bibfnamefont {Q.}~\bibnamefont {Barbedienne}}, \bibinfo
  {author} {\bibfnamefont {A.}~\bibnamefont {Marty}}, \bibinfo {author}
  {\bibfnamefont {G.}~\bibnamefont {Isella}}, \bibinfo {author} {\bibfnamefont
  {L.}~\bibnamefont {Cagnon}}, \bibinfo {author} {\bibfnamefont
  {C.}~\bibnamefont {Vergnaud}}, \bibinfo {author} {\bibfnamefont
  {H.}~\bibnamefont {Jaffrès}}, \bibinfo {author} {\bibfnamefont
  {N.}~\bibnamefont {Reyren}}, \bibinfo {author} {\bibfnamefont {J.-M.}\
  \bibnamefont {George}}, \bibinfo {author} {\bibfnamefont {A.}~\bibnamefont
  {Fert}},\ and\ \bibinfo {author} {\bibfnamefont {M.}~\bibnamefont {Jamet}},\
  }\bibfield  {title} {{\selectlanguage {english}\bibinfo {title} {Observation
  of {Large} {Unidirectional} {Rashba} {Magnetoresistance} in {Ge}(111)}},\
  }\href {https://doi.org/10.1103/PhysRevLett.124.027201} {\bibfield  {journal}
  {\bibinfo  {journal} {Physical Review Letters}\ }\textbf {\bibinfo {volume}
  {124}},\ \bibinfo {pages} {027201} (\bibinfo {year} {2020})}\BibitemShut
  {NoStop}%
\bibitem [{\citenamefont {Choe}\ \emph {et~al.}(2019)\citenamefont {Choe},
  \citenamefont {Jin}, \citenamefont {Kim}, \citenamefont {Choi}, \citenamefont
  {Jo}, \citenamefont {Oh}, \citenamefont {Park}, \citenamefont {Jin},
  \citenamefont {Koo}, \citenamefont {Min}, \citenamefont {Hong}, \citenamefont
  {Lee}, \citenamefont {Baek},\ and\ \citenamefont
  {Yoo}}]{choe_gate-tunable_2019}%
  \BibitemOpen
  \bibfield  {author} {\bibinfo {author} {\bibfnamefont {D.}~\bibnamefont
  {Choe}}, \bibinfo {author} {\bibfnamefont {M.-J.}\ \bibnamefont {Jin}},
  \bibinfo {author} {\bibfnamefont {S.-I.}\ \bibnamefont {Kim}}, \bibinfo
  {author} {\bibfnamefont {H.-J.}\ \bibnamefont {Choi}}, \bibinfo {author}
  {\bibfnamefont {J.}~\bibnamefont {Jo}}, \bibinfo {author} {\bibfnamefont
  {I.}~\bibnamefont {Oh}}, \bibinfo {author} {\bibfnamefont {J.}~\bibnamefont
  {Park}}, \bibinfo {author} {\bibfnamefont {H.}~\bibnamefont {Jin}}, \bibinfo
  {author} {\bibfnamefont {H.~C.}\ \bibnamefont {Koo}}, \bibinfo {author}
  {\bibfnamefont {B.-C.}\ \bibnamefont {Min}}, \bibinfo {author} {\bibfnamefont
  {S.}~\bibnamefont {Hong}}, \bibinfo {author} {\bibfnamefont {H.-W.}\
  \bibnamefont {Lee}}, \bibinfo {author} {\bibfnamefont {S.-H.}\ \bibnamefont
  {Baek}},\ and\ \bibinfo {author} {\bibfnamefont {J.-W.}\ \bibnamefont
  {Yoo}},\ }\bibfield  {title} {{\selectlanguage {english}\bibinfo {title}
  {Gate-tunable giant nonreciprocal charge transport in noncentrosymmetric
  oxide interfaces}},\ }\href {https://doi.org/10.1038/s41467-019-12466-1}
  {\bibfield  {journal} {\bibinfo  {journal} {Nature Communications}\ }\textbf
  {\bibinfo {volume} {10}},\ \bibinfo {pages} {4510} (\bibinfo {year}
  {2019})}\BibitemShut {NoStop}%
\bibitem [{\citenamefont {Vaz}\ \emph {et~al.}(2020)\citenamefont {Vaz},
  \citenamefont {Trier}, \citenamefont {Dyrdał}, \citenamefont {Johansson},
  \citenamefont {Garcia}, \citenamefont {Barthélémy}, \citenamefont {Mertig},
  \citenamefont {Barnaś}, \citenamefont {Fert},\ and\ \citenamefont
  {Bibes}}]{vaz_determining_2020}%
  \BibitemOpen
  \bibfield  {author} {\bibinfo {author} {\bibfnamefont {D.~C.}\ \bibnamefont
  {Vaz}}, \bibinfo {author} {\bibfnamefont {F.}~\bibnamefont {Trier}}, \bibinfo
  {author} {\bibfnamefont {A.}~\bibnamefont {Dyrdał}}, \bibinfo {author}
  {\bibfnamefont {A.}~\bibnamefont {Johansson}}, \bibinfo {author}
  {\bibfnamefont {K.}~\bibnamefont {Garcia}}, \bibinfo {author} {\bibfnamefont
  {A.}~\bibnamefont {Barthélémy}}, \bibinfo {author} {\bibfnamefont
  {I.}~\bibnamefont {Mertig}}, \bibinfo {author} {\bibfnamefont
  {J.}~\bibnamefont {Barnaś}}, \bibinfo {author} {\bibfnamefont
  {A.}~\bibnamefont {Fert}},\ and\ \bibinfo {author} {\bibfnamefont
  {M.}~\bibnamefont {Bibes}},\ }\bibfield  {title} {\bibinfo {title}
  {Determining the {Rashba} parameter from the bilinear magnetoresistance
  response in a two-dimensional electron gas},\ }\href
  {https://doi.org/10.1103/PhysRevMaterials.4.071001} {\bibfield  {journal}
  {\bibinfo  {journal} {Physical Review Materials}\ }\textbf {\bibinfo {volume}
  {4}},\ \bibinfo {pages} {071001} (\bibinfo {year} {2020})}\BibitemShut
  {NoStop}%
\bibitem [{\citenamefont {Lee}\ \emph {et~al.}(2021)\citenamefont {Lee},
  \citenamefont {Harada}, \citenamefont {Trier}, \citenamefont {Marcano},
  \citenamefont {Godel}, \citenamefont {Valencia}, \citenamefont {Tsukazaki},\
  and\ \citenamefont {Bibes}}]{lee_nonreciprocal_2021}%
  \BibitemOpen
  \bibfield  {author} {\bibinfo {author} {\bibfnamefont {J.~H.}\ \bibnamefont
  {Lee}}, \bibinfo {author} {\bibfnamefont {T.}~\bibnamefont {Harada}},
  \bibinfo {author} {\bibfnamefont {F.}~\bibnamefont {Trier}}, \bibinfo
  {author} {\bibfnamefont {L.}~\bibnamefont {Marcano}}, \bibinfo {author}
  {\bibfnamefont {F.}~\bibnamefont {Godel}}, \bibinfo {author} {\bibfnamefont
  {S.}~\bibnamefont {Valencia}}, \bibinfo {author} {\bibfnamefont
  {A.}~\bibnamefont {Tsukazaki}},\ and\ \bibinfo {author} {\bibfnamefont
  {M.}~\bibnamefont {Bibes}},\ }\bibfield  {title} {{\selectlanguage
  {english}\bibinfo {title} {Nonreciprocal {Transport} in a {Rashba}
  {Ferromagnet}, {Delafossite} {PdCoO$_2$}}},\ }\href
  {https://doi.org/10.1021/acs.nanolett.1c02756} {\bibfield  {journal}
  {\bibinfo  {journal} {Nano Letters}\ }\textbf {\bibinfo {volume} {21}},\
  \bibinfo {pages} {8687} (\bibinfo {year} {2021})}\BibitemShut {NoStop}%
\bibitem [{\citenamefont {Vicente‐Arche}\ \emph {et~al.}(2021)\citenamefont
  {Vicente‐Arche}, \citenamefont {Bréhin}, \citenamefont {Varotto},
  \citenamefont {Cosset‐Cheneau}, \citenamefont {Mallik}, \citenamefont
  {Salazar}, \citenamefont {Noël}, \citenamefont {Vaz}, \citenamefont {Trier},
  \citenamefont {Bhattacharya}, \citenamefont {Sander}, \citenamefont
  {Le~Fèvre}, \citenamefont {Bertran}, \citenamefont {Saiz}, \citenamefont
  {Ménard}, \citenamefont {Bergeal}, \citenamefont {Barthélémy},
  \citenamefont {Li}, \citenamefont {Lin}, \citenamefont {Nikonov},
  \citenamefont {Young}, \citenamefont {Rault}, \citenamefont {Vila},
  \citenamefont {Attané},\ and\ \citenamefont
  {Bibes}}]{vicentearche_spincharge_2021}%
  \BibitemOpen
  \bibfield  {author} {\bibinfo {author} {\bibfnamefont {L.~M.}\ \bibnamefont
  {Vicente‐Arche}}, \bibinfo {author} {\bibfnamefont {J.}~\bibnamefont
  {Bréhin}}, \bibinfo {author} {\bibfnamefont {S.}~\bibnamefont {Varotto}},
  \bibinfo {author} {\bibfnamefont {M.}~\bibnamefont {Cosset‐Cheneau}},
  \bibinfo {author} {\bibfnamefont {S.}~\bibnamefont {Mallik}}, \bibinfo
  {author} {\bibfnamefont {R.}~\bibnamefont {Salazar}}, \bibinfo {author}
  {\bibfnamefont {P.}~\bibnamefont {Noël}}, \bibinfo {author} {\bibfnamefont
  {D.~C.}\ \bibnamefont {Vaz}}, \bibinfo {author} {\bibfnamefont
  {F.}~\bibnamefont {Trier}}, \bibinfo {author} {\bibfnamefont
  {S.}~\bibnamefont {Bhattacharya}}, \bibinfo {author} {\bibfnamefont
  {A.}~\bibnamefont {Sander}}, \bibinfo {author} {\bibfnamefont
  {P.}~\bibnamefont {Le~Fèvre}}, \bibinfo {author} {\bibfnamefont
  {F.}~\bibnamefont {Bertran}}, \bibinfo {author} {\bibfnamefont
  {G.}~\bibnamefont {Saiz}}, \bibinfo {author} {\bibfnamefont {G.}~\bibnamefont
  {Ménard}}, \bibinfo {author} {\bibfnamefont {N.}~\bibnamefont {Bergeal}},
  \bibinfo {author} {\bibfnamefont {A.}~\bibnamefont {Barthélémy}}, \bibinfo
  {author} {\bibfnamefont {H.}~\bibnamefont {Li}}, \bibinfo {author}
  {\bibfnamefont {C.}~\bibnamefont {Lin}}, \bibinfo {author} {\bibfnamefont
  {D.~E.}\ \bibnamefont {Nikonov}}, \bibinfo {author} {\bibfnamefont {I.~A.}\
  \bibnamefont {Young}}, \bibinfo {author} {\bibfnamefont {J.~E.}\ \bibnamefont
  {Rault}}, \bibinfo {author} {\bibfnamefont {L.}~\bibnamefont {Vila}},
  \bibinfo {author} {\bibfnamefont {J.}~\bibnamefont {Attané}},\ and\ \bibinfo
  {author} {\bibfnamefont {M.}~\bibnamefont {Bibes}},\ }\bibfield  {title}
  {{\selectlanguage {english}\bibinfo {title} {Spin–{Charge}
  {Interconversion} in {KTaO$_3$}  {2D} {Electron} {Gases}}},\
  }\href {https://onlinelibrary.wiley.com/doi/10.1002/adma.202102102}
  {\bibfield  {journal} {\bibinfo  {journal} {Advanced Materials}\ ,\ \bibinfo
  {pages} {2102102}} (\bibinfo {year} {2021})}\BibitemShut {NoStop}%
\bibitem [{\citenamefont {Vaz}\ \emph {et~al.}(2019)\citenamefont {Vaz},
  \citenamefont {Noël}, \citenamefont {Johansson}, \citenamefont {Göbel},
  \citenamefont {Bruno}, \citenamefont {Singh}, \citenamefont {McKeown-Walker},
  \citenamefont {Trier}, \citenamefont {Vicente-Arche}, \citenamefont {Sander},
  \citenamefont {Valencia}, \citenamefont {Bruneel}, \citenamefont {Vivek},
  \citenamefont {Gabay}, \citenamefont {Bergeal}, \citenamefont {Baumberger},
  \citenamefont {Okuno}, \citenamefont {Barthélémy}, \citenamefont {Fert},
  \citenamefont {Vila}, \citenamefont {Mertig}, \citenamefont {Attané},\ and\
  \citenamefont {Bibes}}]{vaz_mapping_2019}%
  \BibitemOpen
  \bibfield  {author} {\bibinfo {author} {\bibfnamefont {D.~C.}\ \bibnamefont
  {Vaz}}, \bibinfo {author} {\bibfnamefont {P.}~\bibnamefont {Noël}}, \bibinfo
  {author} {\bibfnamefont {A.}~\bibnamefont {Johansson}}, \bibinfo {author}
  {\bibfnamefont {B.}~\bibnamefont {Göbel}}, \bibinfo {author} {\bibfnamefont
  {F.~Y.}\ \bibnamefont {Bruno}}, \bibinfo {author} {\bibfnamefont
  {G.}~\bibnamefont {Singh}}, \bibinfo {author} {\bibfnamefont
  {S.}~\bibnamefont {McKeown-Walker}}, \bibinfo {author} {\bibfnamefont
  {F.}~\bibnamefont {Trier}}, \bibinfo {author} {\bibfnamefont {L.~M.}\
  \bibnamefont {Vicente-Arche}}, \bibinfo {author} {\bibfnamefont
  {A.}~\bibnamefont {Sander}}, \bibinfo {author} {\bibfnamefont
  {S.}~\bibnamefont {Valencia}}, \bibinfo {author} {\bibfnamefont
  {P.}~\bibnamefont {Bruneel}}, \bibinfo {author} {\bibfnamefont
  {M.}~\bibnamefont {Vivek}}, \bibinfo {author} {\bibfnamefont
  {M.}~\bibnamefont {Gabay}}, \bibinfo {author} {\bibfnamefont
  {N.}~\bibnamefont {Bergeal}}, \bibinfo {author} {\bibfnamefont
  {F.}~\bibnamefont {Baumberger}}, \bibinfo {author} {\bibfnamefont
  {H.}~\bibnamefont {Okuno}}, \bibinfo {author} {\bibfnamefont
  {A.}~\bibnamefont {Barthélémy}}, \bibinfo {author} {\bibfnamefont
  {A.}~\bibnamefont {Fert}}, \bibinfo {author} {\bibfnamefont {L.}~\bibnamefont
  {Vila}}, \bibinfo {author} {\bibfnamefont {I.}~\bibnamefont {Mertig}},
  \bibinfo {author} {\bibfnamefont {J.-P.}\ \bibnamefont {Attané}},\ and\
  \bibinfo {author} {\bibfnamefont {M.}~\bibnamefont {Bibes}},\ }\bibfield
  {title} {{\selectlanguage {english}\bibinfo {title} {Mapping spin–charge
  conversion to the band structure in a topological oxide two-dimensional
  electron gas}},\ }\href {https://doi.org/10.1038/s41563-019-0467-4}
  {\bibfield  {journal} {\bibinfo  {journal} {Nature Materials}\ }\textbf
  {\bibinfo {volume} {18}},\ \bibinfo {pages} {1187} (\bibinfo {year}
  {2019})}\BibitemShut {NoStop}%
\bibitem [{\citenamefont {Trier}\ \emph {et~al.}(2020)\citenamefont {Trier},
  \citenamefont {Vaz}, \citenamefont {Bruneel}, \citenamefont {Noël},
  \citenamefont {Fert}, \citenamefont {Vila}, \citenamefont {Attané},
  \citenamefont {Barthélémy}, \citenamefont {Gabay}, \citenamefont
  {Jaffrès},\ and\ \citenamefont {Bibes}}]{trier_electric-field_2020}%
  \BibitemOpen
  \bibfield  {author} {\bibinfo {author} {\bibfnamefont {F.}~\bibnamefont
  {Trier}}, \bibinfo {author} {\bibfnamefont {D.~C.}\ \bibnamefont {Vaz}},
  \bibinfo {author} {\bibfnamefont {P.}~\bibnamefont {Bruneel}}, \bibinfo
  {author} {\bibfnamefont {P.}~\bibnamefont {Noël}}, \bibinfo {author}
  {\bibfnamefont {A.}~\bibnamefont {Fert}}, \bibinfo {author} {\bibfnamefont
  {L.}~\bibnamefont {Vila}}, \bibinfo {author} {\bibfnamefont {J.-P.}\
  \bibnamefont {Attané}}, \bibinfo {author} {\bibfnamefont {A.}~\bibnamefont
  {Barthélémy}}, \bibinfo {author} {\bibfnamefont {M.}~\bibnamefont {Gabay}},
  \bibinfo {author} {\bibfnamefont {H.}~\bibnamefont {Jaffrès}},\ and\
  \bibinfo {author} {\bibfnamefont {M.}~\bibnamefont {Bibes}},\ }\bibfield
  {title} {{\selectlanguage {english}\bibinfo {title} {Electric-{Field}
  {Control} of {Spin} {Current} {Generation} and {Detection} in
  {Ferromagnet}-{Free} {SrTiO$_3$}-{Based} {Nanodevices}}},\
  }\href {https://doi.org/10.1021/acs.nanolett.9b04079} {\bibfield  {journal}
  {\bibinfo  {journal} {Nano Letters}\ }\textbf {\bibinfo {volume} {20}},\
  \bibinfo {pages} {395} (\bibinfo {year} {2020})}\BibitemShut {NoStop}%
\bibitem [{\citenamefont {Bychkov}\ and\ \citenamefont
  {Rashba}(1984)}]{bychkov_properties_1984}%
  \BibitemOpen
  \bibfield  {author} {\bibinfo {author} {\bibfnamefont {Y.}~\bibnamefont
  {Bychkov}, \bibfnamefont {A.}}\ and\ \bibinfo {author} {\bibfnamefont
  {E.~I.}\ \bibnamefont {Rashba}},\ }\bibfield  {title} {\bibinfo {title}
  {Properties of a {2D} electron gas with lifted spectral degeneracy},\
  }\href@noop {} {\bibfield  {journal} {\bibinfo  {journal} {JETP Letters}\
  }\textbf {\bibinfo {volume} {39}},\ \bibinfo {pages} {78} (\bibinfo {year}
  {1984})}\BibitemShut {NoStop}%
\bibitem [{\citenamefont {Manchon}\ \emph {et~al.}(2015)\citenamefont
  {Manchon}, \citenamefont {Koo}, \citenamefont {Nitta}, \citenamefont
  {Frolov},\ and\ \citenamefont {Duine}}]{manchon_new_2015}%
  \BibitemOpen
  \bibfield  {author} {\bibinfo {author} {\bibfnamefont {A.}~\bibnamefont
  {Manchon}}, \bibinfo {author} {\bibfnamefont {H.~C.}\ \bibnamefont {Koo}},
  \bibinfo {author} {\bibfnamefont {J.}~\bibnamefont {Nitta}}, \bibinfo
  {author} {\bibfnamefont {S.~M.}\ \bibnamefont {Frolov}},\ and\ \bibinfo
  {author} {\bibfnamefont {R.~A.}\ \bibnamefont {Duine}},\ }\bibfield  {title}
  {{\selectlanguage {english}\bibinfo {title} {New perspectives for {Rashba}
  spin–orbit coupling}},\ }\href {https://doi.org/10.1038/nmat4360}
  {\bibfield  {journal} {\bibinfo  {journal} {Nature Materials}\ }\textbf
  {\bibinfo {volume} {14}},\ \bibinfo {pages} {871} (\bibinfo {year}
  {2015})}\BibitemShut {NoStop}%
\bibitem [{\citenamefont {Caviglia}\ \emph {et~al.}(2010)\citenamefont
  {Caviglia}, \citenamefont {Gabay}, \citenamefont {Gariglio}, \citenamefont
  {Reyren}, \citenamefont {Cancellieri},\ and\ \citenamefont
  {Triscone}}]{caviglia_tunable_2010}%
  \BibitemOpen
  \bibfield  {author} {\bibinfo {author} {\bibfnamefont {A.~D.}\ \bibnamefont
  {Caviglia}}, \bibinfo {author} {\bibfnamefont {M.}~\bibnamefont {Gabay}},
  \bibinfo {author} {\bibfnamefont {S.}~\bibnamefont {Gariglio}}, \bibinfo
  {author} {\bibfnamefont {N.}~\bibnamefont {Reyren}}, \bibinfo {author}
  {\bibfnamefont {C.}~\bibnamefont {Cancellieri}},\ and\ \bibinfo {author}
  {\bibfnamefont {J.-M.}\ \bibnamefont {Triscone}},\ }\bibfield  {title}
  {\bibinfo {title} {Tunable {Rashba} {Spin}-{Orbit} {Interaction} at {Oxide}
  {Interfaces}},\ }\href {https://doi.org/10.1103/PhysRevLett.104.126803}
  {\bibfield  {journal} {\bibinfo  {journal} {Physical Review Letters}\
  }\textbf {\bibinfo {volume} {104}},\ \bibinfo {pages} {126803} (\bibinfo
  {year} {2010})}\BibitemShut {NoStop}%
\bibitem [{\citenamefont {Lesne}\ \emph {et~al.}(2016)\citenamefont {Lesne},
  \citenamefont {Fu}, \citenamefont {Oyarzun}, \citenamefont {Rojas-Sánchez},
  \citenamefont {Vaz}, \citenamefont {Naganuma},\ and\ \citenamefont
  {Sicoli}}]{lesne_highly_2016}%
  \BibitemOpen
  \bibfield  {author} {\bibinfo {author} {\bibfnamefont {E.}~\bibnamefont
  {Lesne}}, \bibinfo {author} {\bibfnamefont {Y.}~\bibnamefont {Fu}}, \bibinfo
  {author} {\bibfnamefont {S.}~\bibnamefont {Oyarzun}}, \bibinfo {author}
  {\bibfnamefont {J.~C.}\ \bibnamefont {Rojas-Sánchez}}, \bibinfo {author}
  {\bibfnamefont {D.~C.}\ \bibnamefont {Vaz}}, \bibinfo {author} {\bibfnamefont
  {H.}~\bibnamefont {Naganuma}},\ and\ \bibinfo {author} {\bibfnamefont
  {G.}~\bibnamefont {Sicoli}},\ }\bibfield  {title} {{\selectlanguage
  {english}\bibinfo {title} {Highly efficient and tunable spin-to-charge
  conversion through {Rashba} coupling at oxide interfaces}},\ }\href@noop {}
  {\bibfield  {journal} {\bibinfo  {journal} {Nature Materials}\ }\textbf
  {\bibinfo {volume} {15}},\ \bibinfo {pages} {1261} (\bibinfo {year}
  {2016})}\BibitemShut {NoStop}%
\bibitem [{\citenamefont {Boudjada}\ \emph {et~al.}(2019)\citenamefont
  {Boudjada}, \citenamefont {Khait},\ and\ \citenamefont
  {Paramekanti}}]{boudjada_anisotropic_2019}%
  \BibitemOpen
  \bibfield  {author} {\bibinfo {author} {\bibfnamefont {N.}~\bibnamefont
  {Boudjada}}, \bibinfo {author} {\bibfnamefont {I.}~\bibnamefont {Khait}},\
  and\ \bibinfo {author} {\bibfnamefont {A.}~\bibnamefont {Paramekanti}},\
  }\bibfield  {title} {{\selectlanguage {english}\bibinfo {title} {Anisotropic
  magnetoresistance in multiband systems: {Two}-dimensional electron gases and
  polar metals at oxide interfaces}},\ }\href
  {https://doi.org/10.1103/PhysRevB.99.195453} {\bibfield  {journal} {\bibinfo
  {journal} {Physical Review B}\ }\textbf {\bibinfo {volume} {99}},\ \bibinfo
  {pages} {195453} (\bibinfo {year} {2019})}\BibitemShut {NoStop}%
\bibitem [{\citenamefont {Zhang}\ \emph {et~al.}(2019)\citenamefont {Zhang},
  \citenamefont {Yan}, \citenamefont {Zhang}, \citenamefont {Wang},
  \citenamefont {Xiong}, \citenamefont {Zhang}, \citenamefont {Qi},
  \citenamefont {Zhang}, \citenamefont {Han}, \citenamefont {Wu}, \citenamefont
  {Liu}, \citenamefont {Chen}, \citenamefont {Shen},\ and\ \citenamefont
  {Sun}}]{zhang_unusual_2019}%
  \BibitemOpen
  \bibfield  {author} {\bibinfo {author} {\bibfnamefont {H.}~\bibnamefont
  {Zhang}}, \bibinfo {author} {\bibfnamefont {X.}~\bibnamefont {Yan}}, \bibinfo
  {author} {\bibfnamefont {X.}~\bibnamefont {Zhang}}, \bibinfo {author}
  {\bibfnamefont {S.}~\bibnamefont {Wang}}, \bibinfo {author} {\bibfnamefont
  {C.}~\bibnamefont {Xiong}}, \bibinfo {author} {\bibfnamefont
  {H.}~\bibnamefont {Zhang}}, \bibinfo {author} {\bibfnamefont
  {S.}~\bibnamefont {Qi}}, \bibinfo {author} {\bibfnamefont {J.}~\bibnamefont
  {Zhang}}, \bibinfo {author} {\bibfnamefont {F.}~\bibnamefont {Han}}, \bibinfo
  {author} {\bibfnamefont {N.}~\bibnamefont {Wu}}, \bibinfo {author}
  {\bibfnamefont {B.}~\bibnamefont {Liu}}, \bibinfo {author} {\bibfnamefont
  {Y.}~\bibnamefont {Chen}}, \bibinfo {author} {\bibfnamefont {B.}~\bibnamefont
  {Shen}},\ and\ \bibinfo {author} {\bibfnamefont {J.}~\bibnamefont {Sun}},\
  }\bibfield  {title} {{\selectlanguage {english}\bibinfo {title} {Unusual
  {Electric} and {Optical} {Tuning} of {KTaO$_3$}-{Based}
  {Two}-{Dimensional} {Electron} {Gases} with 5d {Orbitals}}},\ }\href
  {https://doi.org/10.1021/acsnano.8b07622} {\bibfield  {journal} {\bibinfo
  {journal} {ACS Nano}\ }\textbf {\bibinfo {volume} {13}},\ \bibinfo {pages}
  {609} (\bibinfo {year} {2019})}\BibitemShut {NoStop}%
\bibitem [{\citenamefont {Varotto}\ \emph {et~al.}(2022)\citenamefont
  {Varotto}, \citenamefont {Johansson}, \citenamefont {Göbel}, \citenamefont
  {Vicente-Arche}, \citenamefont {Mallik}, \citenamefont {Bréhin},
  \citenamefont {Salazar}, \citenamefont {Bertran}, \citenamefont {Fèvre},
  \citenamefont {Bergeal}, \citenamefont {Rault}, \citenamefont {Mertig},\ and\
  \citenamefont {Bibes}}]{varotto_direct_2022}%
  \BibitemOpen
  \bibfield  {author} {\bibinfo {author} {\bibfnamefont {S.}~\bibnamefont
  {Varotto}}, \bibinfo {author} {\bibfnamefont {A.}~\bibnamefont {Johansson}},
  \bibinfo {author} {\bibfnamefont {B.}~\bibnamefont {Göbel}}, \bibinfo
  {author} {\bibfnamefont {L.~M.}\ \bibnamefont {Vicente-Arche}}, \bibinfo
  {author} {\bibfnamefont {S.}~\bibnamefont {Mallik}}, \bibinfo {author}
  {\bibfnamefont {J.}~\bibnamefont {Bréhin}}, \bibinfo {author} {\bibfnamefont
  {R.}~\bibnamefont {Salazar}}, \bibinfo {author} {\bibfnamefont
  {F.}~\bibnamefont {Bertran}}, \bibinfo {author} {\bibfnamefont {P.~L.}\
  \bibnamefont {Fèvre}}, \bibinfo {author} {\bibfnamefont {N.}~\bibnamefont
  {Bergeal}}, \bibinfo {author} {\bibfnamefont {J.}~\bibnamefont {Rault}},
  \bibinfo {author} {\bibfnamefont {I.}~\bibnamefont {Mertig}},\ and\ \bibinfo
  {author} {\bibfnamefont {M.}~\bibnamefont {Bibes}},\ }\bibfield  {title}
  {{\selectlanguage {english}\bibinfo {title} {Direct visualization of
  {Rashba}-split bands and spin/orbital-charge interconversion at {KTaO$_3$}
  interfaces}},\ }\href {https://doi.org/10.1038/s41467-022-33621-1} {\bibfield
   {journal} {\bibinfo  {journal} {Nature Communications}\ }\textbf {\bibinfo
  {volume} {13}},\ \bibinfo {pages} {6165} (\bibinfo {year}
  {2022})}\BibitemShut {NoStop}%
\bibitem [{\citenamefont {Vicente-Arche}\ \emph {et~al.}(2021)\citenamefont
  {Vicente-Arche}, \citenamefont {Mallik}, \citenamefont {Cosset-Cheneau},
  \citenamefont {Noël}, \citenamefont {Vaz}, \citenamefont {Trier},
  \citenamefont {Gosavi}, \citenamefont {Lin}, \citenamefont {Nikonov},
  \citenamefont {Young}, \citenamefont {Sander}, \citenamefont {Barthélémy},
  \citenamefont {Attané}, \citenamefont {Vila},\ and\ \citenamefont
  {Bibes}}]{vicente-arche_metal_2021}%
  \BibitemOpen
  \bibfield  {author} {\bibinfo {author} {\bibfnamefont {L.~M.}\ \bibnamefont
  {Vicente-Arche}}, \bibinfo {author} {\bibfnamefont {S.}~\bibnamefont
  {Mallik}}, \bibinfo {author} {\bibfnamefont {M.}~\bibnamefont
  {Cosset-Cheneau}}, \bibinfo {author} {\bibfnamefont {P.}~\bibnamefont
  {Noël}}, \bibinfo {author} {\bibfnamefont {D.~C.}\ \bibnamefont {Vaz}},
  \bibinfo {author} {\bibfnamefont {F.}~\bibnamefont {Trier}}, \bibinfo
  {author} {\bibfnamefont {T.~A.}\ \bibnamefont {Gosavi}}, \bibinfo {author}
  {\bibfnamefont {C.-C.}\ \bibnamefont {Lin}}, \bibinfo {author} {\bibfnamefont
  {D.~E.}\ \bibnamefont {Nikonov}}, \bibinfo {author} {\bibfnamefont {I.~A.}\
  \bibnamefont {Young}}, \bibinfo {author} {\bibfnamefont {A.}~\bibnamefont
  {Sander}}, \bibinfo {author} {\bibfnamefont {A.}~\bibnamefont
  {Barthélémy}}, \bibinfo {author} {\bibfnamefont {J.-P.}\ \bibnamefont
  {Attané}}, \bibinfo {author} {\bibfnamefont {L.}~\bibnamefont {Vila}},\ and\
  \bibinfo {author} {\bibfnamefont {M.}~\bibnamefont {Bibes}},\ }\bibfield
  {title} {{\selectlanguage {english}\bibinfo {title} {Metal/{SrTiO$_3$}
  two-dimensional electron gases for spin-to-charge conversion}},\ }\href
  {https://doi.org/10.1103/PhysRevMaterials.5.064005} {\bibfield  {journal}
  {\bibinfo  {journal} {Physical Review Materials}\ }\textbf {\bibinfo {volume}
  {5}},\ \bibinfo {pages} {064005} (\bibinfo {year} {2021})}\BibitemShut
  {NoStop}%
\bibitem [{sup()}]{supplemental_material}%
  \BibitemOpen
  \href@noop {} {\bibinfo {title} {Supplemental material}},\ \bibinfo
  {howpublished} {\url{URL_will_be_inserted_by_publisher}}\BibitemShut
  {NoStop}%
\bibitem [{\citenamefont {Johansson}\ \emph {et~al.}(2021)\citenamefont
  {Johansson}, \citenamefont {Göbel}, \citenamefont {Henk}, \citenamefont
  {Bibes},\ and\ \citenamefont {Mertig}}]{johansson_spin_2021}%
  \BibitemOpen
  \bibfield  {author} {\bibinfo {author} {\bibfnamefont {A.}~\bibnamefont
  {Johansson}}, \bibinfo {author} {\bibfnamefont {B.}~\bibnamefont {Göbel}},
  \bibinfo {author} {\bibfnamefont {J.}~\bibnamefont {Henk}}, \bibinfo {author}
  {\bibfnamefont {M.}~\bibnamefont {Bibes}},\ and\ \bibinfo {author}
  {\bibfnamefont {I.}~\bibnamefont {Mertig}},\ }\bibfield  {title}
  {{\selectlanguage {english}\bibinfo {title} {Spin and orbital {Edelstein}
  effects in a two-dimensional electron gas: {Theory} and application to
  {SrTiO$_3$} interfaces}},\ }\href
  {https://doi.org/10.1103/PhysRevResearch.3.013275} {\bibfield  {journal}
  {\bibinfo  {journal} {Physical Review Research}\ }\textbf {\bibinfo {volume}
  {3}},\ \bibinfo {pages} {013275} (\bibinfo {year} {2021})}\BibitemShut
  {NoStop}%
\bibitem [{\citenamefont {Hurand}\ \emph {et~al.}(2015)\citenamefont {Hurand},
  \citenamefont {Jouan}, \citenamefont {Feuillet-Palma}, \citenamefont {Singh},
  \citenamefont {Biscaras}, \citenamefont {Lesne}, \citenamefont {Reyren},
  \citenamefont {Barthélémy}, \citenamefont {Bibes}, \citenamefont
  {Villegas}, \citenamefont {Ulysse}, \citenamefont {Lafosse}, \citenamefont
  {Pannetier-Lecoeur}, \citenamefont {Caprara}, \citenamefont {Grilli},
  \citenamefont {Lesueur},\ and\ \citenamefont
  {Bergeal}}]{hurand_field-effect_2015}%
  \BibitemOpen
  \bibfield  {author} {\bibinfo {author} {\bibfnamefont {S.}~\bibnamefont
  {Hurand}}, \bibinfo {author} {\bibfnamefont {A.}~\bibnamefont {Jouan}},
  \bibinfo {author} {\bibfnamefont {C.}~\bibnamefont {Feuillet-Palma}},
  \bibinfo {author} {\bibfnamefont {G.}~\bibnamefont {Singh}}, \bibinfo
  {author} {\bibfnamefont {J.}~\bibnamefont {Biscaras}}, \bibinfo {author}
  {\bibfnamefont {E.}~\bibnamefont {Lesne}}, \bibinfo {author} {\bibfnamefont
  {N.}~\bibnamefont {Reyren}}, \bibinfo {author} {\bibfnamefont
  {A.}~\bibnamefont {Barthélémy}}, \bibinfo {author} {\bibfnamefont
  {M.}~\bibnamefont {Bibes}}, \bibinfo {author} {\bibfnamefont {J.~E.}\
  \bibnamefont {Villegas}}, \bibinfo {author} {\bibfnamefont {C.}~\bibnamefont
  {Ulysse}}, \bibinfo {author} {\bibfnamefont {X.}~\bibnamefont {Lafosse}},
  \bibinfo {author} {\bibfnamefont {M.}~\bibnamefont {Pannetier-Lecoeur}},
  \bibinfo {author} {\bibfnamefont {S.}~\bibnamefont {Caprara}}, \bibinfo
  {author} {\bibfnamefont {M.}~\bibnamefont {Grilli}}, \bibinfo {author}
  {\bibfnamefont {J.}~\bibnamefont {Lesueur}},\ and\ \bibinfo {author}
  {\bibfnamefont {N.}~\bibnamefont {Bergeal}},\ }\bibfield  {title}
  {{\selectlanguage {english}\bibinfo {title} {Field-effect control of
  superconductivity and {Rashba} spin-orbit coupling in top-gated
  {LaAlO$_3$}/{SrTiO$_3$} devices}},\ }\href {https://doi.org/10.1038/srep12751}
  {\bibfield  {journal} {\bibinfo  {journal} {Scientific Reports}\ }\textbf
  {\bibinfo {volume} {5}},\ \bibinfo {pages} {12751} (\bibinfo {year}
  {2015})}\BibitemShut {NoStop}%
\bibitem [{\citenamefont {Jouan}\ \emph {et~al.}(2022)\citenamefont {Jouan},
  \citenamefont {Hurand}, \citenamefont {Singh}, \citenamefont {Lesne},
  \citenamefont {Barthélémy}, \citenamefont {Bibes}, \citenamefont {Ulysse},
  \citenamefont {Saiz}, \citenamefont {Feuillet‐Palma}, \citenamefont
  {Lesueur},\ and\ \citenamefont {Bergeal}}]{jouan_multiband_2022}%
  \BibitemOpen
  \bibfield  {author} {\bibinfo {author} {\bibfnamefont {A.}~\bibnamefont
  {Jouan}}, \bibinfo {author} {\bibfnamefont {S.}~\bibnamefont {Hurand}},
  \bibinfo {author} {\bibfnamefont {G.}~\bibnamefont {Singh}}, \bibinfo
  {author} {\bibfnamefont {E.}~\bibnamefont {Lesne}}, \bibinfo {author}
  {\bibfnamefont {A.}~\bibnamefont {Barthélémy}}, \bibinfo {author}
  {\bibfnamefont {M.}~\bibnamefont {Bibes}}, \bibinfo {author} {\bibfnamefont
  {C.}~\bibnamefont {Ulysse}}, \bibinfo {author} {\bibfnamefont
  {G.}~\bibnamefont {Saiz}}, \bibinfo {author} {\bibfnamefont {C.}~\bibnamefont
  {Feuillet‐Palma}}, \bibinfo {author} {\bibfnamefont {J.}~\bibnamefont
  {Lesueur}},\ and\ \bibinfo {author} {\bibfnamefont {N.}~\bibnamefont
  {Bergeal}},\ }\bibfield  {title} {{\selectlanguage {english}\bibinfo {title}
  {Multiband {Effects} in the {Superconducting} {Phase} {Diagram} of {Oxide}
  {Interfaces}}},\ }\href {https://doi.org/10.1002/admi.202201392} {\bibfield
  {journal} {\bibinfo  {journal} {Advanced Materials Interfaces}\ }\textbf
  {\bibinfo {volume} {9}},\ \bibinfo {pages} {2201392} (\bibinfo {year}
  {2022})}\BibitemShut {NoStop}%
\bibitem [{\citenamefont {Tokura}\ and\ \citenamefont
  {Nagaosa}(2018)}]{tokura_nonreciprocal_2018}%
  \BibitemOpen
  \bibfield  {author} {\bibinfo {author} {\bibfnamefont {Y.}~\bibnamefont
  {Tokura}}\ and\ \bibinfo {author} {\bibfnamefont {N.}~\bibnamefont
  {Nagaosa}},\ }\bibfield  {title} {{\selectlanguage {english}\bibinfo {title}
  {Nonreciprocal responses from non-centrosymmetric quantum materials}},\
  }\href {https://doi.org/10.1038/s41467-018-05759-4} {\bibfield  {journal}
  {\bibinfo  {journal} {Nature Communications}\ }\textbf {\bibinfo {volume}
  {9}},\ \bibinfo {pages} {3740} (\bibinfo {year} {2018})}\BibitemShut
  {NoStop}%
\bibitem [{\citenamefont {Fu}\ \emph {et~al.}(2022)\citenamefont {Fu},
  \citenamefont {Li}, \citenamefont {Papin}, \citenamefont {Noël},
  \citenamefont {Teresi}, \citenamefont {Cosset-Chéneau}, \citenamefont
  {Grezes}, \citenamefont {Guillet}, \citenamefont {Thomas}, \citenamefont
  {Niquet}, \citenamefont {Ballet}, \citenamefont {Meunier}, \citenamefont
  {Attané}, \citenamefont {Fert},\ and\ \citenamefont
  {Vila}}]{fu_bilinear_2022}%
  \BibitemOpen
  \bibfield  {author} {\bibinfo {author} {\bibfnamefont {Y.}~\bibnamefont
  {Fu}}, \bibinfo {author} {\bibfnamefont {J.}~\bibnamefont {Li}}, \bibinfo
  {author} {\bibfnamefont {J.}~\bibnamefont {Papin}}, \bibinfo {author}
  {\bibfnamefont {P.}~\bibnamefont {Noël}}, \bibinfo {author} {\bibfnamefont
  {S.}~\bibnamefont {Teresi}}, \bibinfo {author} {\bibfnamefont
  {M.}~\bibnamefont {Cosset-Chéneau}}, \bibinfo {author} {\bibfnamefont
  {C.}~\bibnamefont {Grezes}}, \bibinfo {author} {\bibfnamefont
  {T.}~\bibnamefont {Guillet}}, \bibinfo {author} {\bibfnamefont
  {C.}~\bibnamefont {Thomas}}, \bibinfo {author} {\bibfnamefont {Y.-M.}\
  \bibnamefont {Niquet}}, \bibinfo {author} {\bibfnamefont {P.}~\bibnamefont
  {Ballet}}, \bibinfo {author} {\bibfnamefont {T.}~\bibnamefont {Meunier}},
  \bibinfo {author} {\bibfnamefont {J.-P.}\ \bibnamefont {Attané}}, \bibinfo
  {author} {\bibfnamefont {A.}~\bibnamefont {Fert}},\ and\ \bibinfo {author}
  {\bibfnamefont {L.}~\bibnamefont {Vila}},\ }\bibfield  {title}
  {{\selectlanguage {english}\bibinfo {title} {Bilinear {Magnetoresistance} in
  {HgTe} {Topological} {Insulator}: {Opposite} {Signs} at {Opposite} {Surfaces}
  {Demonstrated} by {Gate} {Control}}},\ }\href
  {https://doi.org/10.1021/acs.nanolett.2c02585} {\bibfield  {journal}
  {\bibinfo  {journal} {Nano Letters}\ }\textbf {\bibinfo {volume} {22}},\
  \bibinfo {pages} {7867} (\bibinfo {year} {2022})}\BibitemShut {NoStop}%
\bibitem [{\citenamefont {Yasuda}\ \emph {et~al.}(2016)\citenamefont {Yasuda},
  \citenamefont {Tsukazaki}, \citenamefont {Yoshimi}, \citenamefont
  {Takahashi}, \citenamefont {Kawasaki},\ and\ \citenamefont
  {Tokura}}]{yasuda_large_2016}%
  \BibitemOpen
  \bibfield  {author} {\bibinfo {author} {\bibfnamefont {K.}~\bibnamefont
  {Yasuda}}, \bibinfo {author} {\bibfnamefont {A.}~\bibnamefont {Tsukazaki}},
  \bibinfo {author} {\bibfnamefont {R.}~\bibnamefont {Yoshimi}}, \bibinfo
  {author} {\bibfnamefont {K.~S.}\ \bibnamefont {Takahashi}}, \bibinfo {author}
  {\bibfnamefont {M.}~\bibnamefont {Kawasaki}},\ and\ \bibinfo {author}
  {\bibfnamefont {Y.}~\bibnamefont {Tokura}},\ }\bibfield  {title}
  {{\selectlanguage {english}\bibinfo {title} {Large {Unidirectional}
  {Magnetoresistance} in a {Magnetic} {Topological} {Insulator}}},\ }\href@noop
  {} {\bibfield  {journal} {\bibinfo  {journal} {Physical Review Letters}\
  }\textbf {\bibinfo {volume} {117}},\ \bibinfo {pages} {127202} (\bibinfo
  {year} {2016})}\BibitemShut {NoStop}%
\bibitem [{\citenamefont {Zhang}\ \emph {et~al.}(2022)\citenamefont {Zhang},
  \citenamefont {Wang}, \citenamefont {Cao}, \citenamefont {Wang},
  \citenamefont {Zhou}, \citenamefont {Watanabe}, \citenamefont {Taniguchi},
  \citenamefont {Yan},\ and\ \citenamefont {Gao}}]{zhang_controlled_2022}%
  \BibitemOpen
  \bibfield  {author} {\bibinfo {author} {\bibfnamefont {Z.}~\bibnamefont
  {Zhang}}, \bibinfo {author} {\bibfnamefont {N.}~\bibnamefont {Wang}},
  \bibinfo {author} {\bibfnamefont {N.}~\bibnamefont {Cao}}, \bibinfo {author}
  {\bibfnamefont {A.}~\bibnamefont {Wang}}, \bibinfo {author} {\bibfnamefont
  {X.}~\bibnamefont {Zhou}}, \bibinfo {author} {\bibfnamefont {K.}~\bibnamefont
  {Watanabe}}, \bibinfo {author} {\bibfnamefont {T.}~\bibnamefont {Taniguchi}},
  \bibinfo {author} {\bibfnamefont {B.}~\bibnamefont {Yan}},\ and\ \bibinfo
  {author} {\bibfnamefont {W.-b.}\ \bibnamefont {Gao}},\ }\bibfield  {title}
  {{\selectlanguage {english}\bibinfo {title} {Controlled large non-reciprocal
  charge transport in an intrinsic magnetic topological insulator
  {MnBi$_2$Te$_4$}}},\ }\href {https://doi.org/10.1038/s41467-022-33705-y} {\bibfield
   {journal} {\bibinfo  {journal} {Nature Communications}\ }\textbf {\bibinfo
  {volume} {13}},\ \bibinfo {pages} {6191} (\bibinfo {year}
  {2022})}\BibitemShut {NoStop}%
\bibitem [{\citenamefont {Wakatsuki}\ \emph {et~al.}(2017)\citenamefont
  {Wakatsuki}, \citenamefont {Saito}, \citenamefont {Hoshino}, \citenamefont
  {Itahashi}, \citenamefont {Ideue}, \citenamefont {Ezawa}, \citenamefont
  {Iwasa},\ and\ \citenamefont {Nagaosa}}]{wakatsuki_nonreciprocal_2017}%
  \BibitemOpen
  \bibfield  {author} {\bibinfo {author} {\bibfnamefont {R.}~\bibnamefont
  {Wakatsuki}}, \bibinfo {author} {\bibfnamefont {Y.}~\bibnamefont {Saito}},
  \bibinfo {author} {\bibfnamefont {S.}~\bibnamefont {Hoshino}}, \bibinfo
  {author} {\bibfnamefont {Y.~M.}\ \bibnamefont {Itahashi}}, \bibinfo {author}
  {\bibfnamefont {T.}~\bibnamefont {Ideue}}, \bibinfo {author} {\bibfnamefont
  {M.}~\bibnamefont {Ezawa}}, \bibinfo {author} {\bibfnamefont
  {Y.}~\bibnamefont {Iwasa}},\ and\ \bibinfo {author} {\bibfnamefont
  {N.}~\bibnamefont {Nagaosa}},\ }\bibfield  {title} {{\selectlanguage
  {english}\bibinfo {title} {Nonreciprocal charge transport in
  noncentrosymmetric superconductors}},\ }\href
  {https://doi.org/10.1126/sciadv.1602390} {\bibfield  {journal} {\bibinfo
  {journal} {Science Advances}\ }\textbf {\bibinfo {volume} {3}},\ \bibinfo
  {pages} {e1602390} (\bibinfo {year} {2017})}\BibitemShut {NoStop}%
\bibitem [{\citenamefont {Li}\ \emph {et~al.}(2021)\citenamefont {Li},
  \citenamefont {Li}, \citenamefont {Li}, \citenamefont {Fang}, \citenamefont
  {Yang}, \citenamefont {Wen}, \citenamefont {Zheng}, \citenamefont {Zhang},
  \citenamefont {He}, \citenamefont {Manchon}, \citenamefont {Cheng},\ and\
  \citenamefont {Zhang}}]{li_nonreciprocal_2021}%
  \BibitemOpen
  \bibfield  {author} {\bibinfo {author} {\bibfnamefont {Y.}~\bibnamefont
  {Li}}, \bibinfo {author} {\bibfnamefont {Y.}~\bibnamefont {Li}}, \bibinfo
  {author} {\bibfnamefont {P.}~\bibnamefont {Li}}, \bibinfo {author}
  {\bibfnamefont {B.}~\bibnamefont {Fang}}, \bibinfo {author} {\bibfnamefont
  {X.}~\bibnamefont {Yang}}, \bibinfo {author} {\bibfnamefont {Y.}~\bibnamefont
  {Wen}}, \bibinfo {author} {\bibfnamefont {D.-x.}\ \bibnamefont {Zheng}},
  \bibinfo {author} {\bibfnamefont {C.-h.}\ \bibnamefont {Zhang}}, \bibinfo
  {author} {\bibfnamefont {X.}~\bibnamefont {He}}, \bibinfo {author}
  {\bibfnamefont {A.}~\bibnamefont {Manchon}}, \bibinfo {author} {\bibfnamefont
  {Z.-H.}\ \bibnamefont {Cheng}},\ and\ \bibinfo {author} {\bibfnamefont
  {X.-x.}\ \bibnamefont {Zhang}},\ }\bibfield  {title} {{\selectlanguage
  {english}\bibinfo {title} {Nonreciprocal charge transport up to room
  temperature in bulk {Rashba} semiconductor $\alpha$-{GeTe}}},\ }\href
  {https://doi.org/10.1038/s41467-020-20840-7} {\bibfield  {journal} {\bibinfo
  {journal} {Nature Communications}\ }\textbf {\bibinfo {volume} {12}},\
  \bibinfo {pages} {540} (\bibinfo {year} {2021})}\BibitemShut {NoStop}%
\bibitem [{\citenamefont {Wang}\ \emph {et~al.}(2017)\citenamefont {Wang},
  \citenamefont {Ramaswamy}, \citenamefont {Motapothula}, \citenamefont
  {Narayanapillai}, \citenamefont {Zhu}, \citenamefont {Yu}, \citenamefont
  {Venkatesan},\ and\ \citenamefont
  {Yang}}]{wangRoomTemperatureGiantChargetoSpin2017}%
  \BibitemOpen
  \bibfield  {author} {\bibinfo {author} {\bibfnamefont {Y.}~\bibnamefont
  {Wang}}, \bibinfo {author} {\bibfnamefont {R.}~\bibnamefont {Ramaswamy}},
  \bibinfo {author} {\bibfnamefont {M.}~\bibnamefont {Motapothula}}, \bibinfo
  {author} {\bibfnamefont {K.}~\bibnamefont {Narayanapillai}}, \bibinfo
  {author} {\bibfnamefont {D.}~\bibnamefont {Zhu}}, \bibinfo {author}
  {\bibfnamefont {J.}~\bibnamefont {Yu}}, \bibinfo {author} {\bibfnamefont
  {T.}~\bibnamefont {Venkatesan}},\ and\ \bibinfo {author} {\bibfnamefont
  {H.}~\bibnamefont {Yang}},\ }\bibfield  {title} {\bibinfo {title}
  {Room-{{Temperature Giant Charge-to-Spin Conversion}} at the {{SrTiO$_3$}}--{{LaAlO$_3$}} {{Oxide
  Interface}}},\ }\href {https://doi.org/10.1021/acs.nanolett.7b03714}
  {\bibfield  {journal} {\bibinfo  {journal} {Nano Letters}\ }\textbf {\bibinfo
  {volume} {17}},\ \bibinfo {pages} {7659} (\bibinfo {year}
  {2017})}\BibitemShut {NoStop}%
\bibitem [{\citenamefont {Shashank}\ \emph {et~al.}(2023)\citenamefont
  {Shashank}, \citenamefont {Deka}, \citenamefont {Ye}, \citenamefont {Gupta},
  \citenamefont {Medwal}, \citenamefont {Rawat}, \citenamefont {Asada},
  \citenamefont {Renshaw~Wang},\ and\ \citenamefont
  {Fukuma}}]{shashankRoomTemperatureCharge2023}%
  \BibitemOpen
  \bibfield  {author} {\bibinfo {author} {\bibfnamefont {U.}~\bibnamefont
  {Shashank}}, \bibinfo {author} {\bibfnamefont {A.}~\bibnamefont {Deka}},
  \bibinfo {author} {\bibfnamefont {C.}~\bibnamefont {Ye}}, \bibinfo {author}
  {\bibfnamefont {S.}~\bibnamefont {Gupta}}, \bibinfo {author} {\bibfnamefont
  {R.}~\bibnamefont {Medwal}}, \bibinfo {author} {\bibfnamefont {R.~S.}\
  \bibnamefont {Rawat}}, \bibinfo {author} {\bibfnamefont {H.}~\bibnamefont
  {Asada}}, \bibinfo {author} {\bibfnamefont {X.}~\bibnamefont
  {Renshaw~Wang}},\ and\ \bibinfo {author} {\bibfnamefont {Y.}~\bibnamefont
  {Fukuma}},\ }\bibfield  {title} {\bibinfo {title} {Room-{{Temperature
  Charge}}-to-{{Spin Conversion}} from {{Quasi}}-{{2D Electron Gas}} at
  {{SrTiO$_3$}}-{{Based Interfaces}}},\ }\href
  {https://doi.org/10.1002/pssr.202200377} {\bibfield  {journal} {\bibinfo
  {journal} {physica status solidi (RRL) \textendash{} Rapid Research Letters}\
  }\textbf {\bibinfo {volume} {17}},\ \bibinfo {pages} {2200377} (\bibinfo
  {year} {2023})}\BibitemShut {NoStop}%
\bibitem [{\citenamefont
  {Varignon}(2023)}]{varignonUnexpectedAntagonismFerroelectricity2023}%
  \BibitemOpen
  \bibfield  {author} {\bibinfo {author} {\bibfnamefont {J.}~\bibnamefont
  {Varignon}},\ }\href@noop {} {\bibinfo {title} {Unexpected antagonism between
  ferroelectricity and {{Rashba}} effects in epitaxially strained
  {{SrTiO$_3$}}}} (\bibinfo {year} {2023}),\ \Eprint
  {https://arxiv.org/abs/2306.12267} {arxiv:2306.12267 [cond-mat]} \BibitemShut
  {NoStop}%
\bibitem [{\citenamefont {Caviglia}\ \emph {et~al.}(2008)\citenamefont
  {Caviglia}, \citenamefont {Gariglio}, \citenamefont {Reyren}, \citenamefont
  {Jaccard}, \citenamefont {Schneider}, \citenamefont {Gabay}, \citenamefont
  {Thiel}, \citenamefont {Hammerl}, \citenamefont {Mannhart},\ and\
  \citenamefont {Triscone}}]{cavigliaElectricFieldControl2008}%
  \BibitemOpen
  \bibfield  {author} {\bibinfo {author} {\bibfnamefont {A.~D.}\ \bibnamefont
  {Caviglia}}, \bibinfo {author} {\bibfnamefont {S.}~\bibnamefont {Gariglio}},
  \bibinfo {author} {\bibfnamefont {N.}~\bibnamefont {Reyren}}, \bibinfo
  {author} {\bibfnamefont {D.}~\bibnamefont {Jaccard}}, \bibinfo {author}
  {\bibfnamefont {T.}~\bibnamefont {Schneider}}, \bibinfo {author}
  {\bibfnamefont {M.}~\bibnamefont {Gabay}}, \bibinfo {author} {\bibfnamefont
  {S.}~\bibnamefont {Thiel}}, \bibinfo {author} {\bibfnamefont
  {G.}~\bibnamefont {Hammerl}}, \bibinfo {author} {\bibfnamefont
  {J.}~\bibnamefont {Mannhart}},\ and\ \bibinfo {author} {\bibfnamefont
  {J.-M.}\ \bibnamefont {Triscone}},\ }\bibfield  {title} {\bibinfo {title}
  {Electric field control of the {{LaAlO$_3$}}/{{SrTiO$_3$}} interface ground
  state},\ }\href {https://doi.org/10.1038/nature07576} {\bibfield  {journal}
  {\bibinfo  {journal} {Nature}\ }\textbf {\bibinfo {volume} {456}},\ \bibinfo
  {pages} {624} (\bibinfo {year} {2008})}\BibitemShut {NoStop}%
\bibitem [{\citenamefont {Joshua}\ \emph {et~al.}(2012)\citenamefont {Joshua},
  \citenamefont {Pecker}, \citenamefont {Ruhman}, \citenamefont {Altman},\ and\
  \citenamefont {Ilani}}]{joshuaUniversalCriticalDensity2012}%
  \BibitemOpen
  \bibfield  {author} {\bibinfo {author} {\bibfnamefont {A.}~\bibnamefont
  {Joshua}}, \bibinfo {author} {\bibfnamefont {S.}~\bibnamefont {Pecker}},
  \bibinfo {author} {\bibfnamefont {J.}~\bibnamefont {Ruhman}}, \bibinfo
  {author} {\bibfnamefont {E.}~\bibnamefont {Altman}},\ and\ \bibinfo {author}
  {\bibfnamefont {S.}~\bibnamefont {Ilani}},\ }\bibfield  {title} {\bibinfo
  {title} {A universal critical density underlying the physics of electrons at
  the {{LaAlO$_3$}}/{{SrTiO$_3$}} interface},\ }\href
  {https://doi.org/10.1038/ncomms2116} {\bibfield  {journal} {\bibinfo
  {journal} {Nature Communications}\ }\textbf {\bibinfo {volume} {3}},\
  \bibinfo {pages} {1129} (\bibinfo {year} {2012})}\BibitemShut {NoStop}%
\bibitem [{\citenamefont {Biscaras}\ \emph {et~al.}(2015)\citenamefont
  {Biscaras}, \citenamefont {Hurand}, \citenamefont {{Feuillet-Palma}},
  \citenamefont {Rastogi}, \citenamefont {Budhani}, \citenamefont {Reyren},
  \citenamefont {Lesne}, \citenamefont {Lesueur},\ and\ \citenamefont
  {Bergeal}}]{biscarasLimitElectrostaticDoping2015}%
  \BibitemOpen
  \bibfield  {author} {\bibinfo {author} {\bibfnamefont {J.}~\bibnamefont
  {Biscaras}}, \bibinfo {author} {\bibfnamefont {S.}~\bibnamefont {Hurand}},
  \bibinfo {author} {\bibfnamefont {C.}~\bibnamefont {{Feuillet-Palma}}},
  \bibinfo {author} {\bibfnamefont {A.}~\bibnamefont {Rastogi}}, \bibinfo
  {author} {\bibfnamefont {R.~C.}\ \bibnamefont {Budhani}}, \bibinfo {author}
  {\bibfnamefont {N.}~\bibnamefont {Reyren}}, \bibinfo {author} {\bibfnamefont
  {E.}~\bibnamefont {Lesne}}, \bibinfo {author} {\bibfnamefont
  {J.}~\bibnamefont {Lesueur}},\ and\ \bibinfo {author} {\bibfnamefont
  {N.}~\bibnamefont {Bergeal}},\ }\bibfield  {title} {\bibinfo {title} {Limit
  of the electrostatic doping in two-dimensional electron gases of
  {{LaXO$_3$}}({{X}} = {{Al}}, {{Ti}})/{{SrTiO$_3$}}},\ }\href
  {https://doi.org/10.1038/srep06788} {\bibfield  {journal} {\bibinfo
  {journal} {Scientific Reports}\ }\textbf {\bibinfo {volume} {4}},\ \bibinfo
  {pages} {6788} (\bibinfo {year} {2015})}\BibitemShut {NoStop}%
\end{thebibliography}

%

\end{document}